\documentclass{article}
\usepackage{jheppub}

\usepackage[utf8]{inputenc}
\usepackage{amssymb}
\usepackage{mathtools}
\usepackage{graphicx}
\usepackage{float}
\usepackage{dsfont} 
\usepackage{cancel}
\usepackage{enumitem}
\usepackage{simplewick}
\usepackage{bbm}
\usepackage{braket}
\usepackage{subcaption}
\usepackage{booktabs, xcolor, braket}
\definecolor{darkgreen}{rgb}{0,0.5,0}
\newcommand{\clr}[1]{\textcolor{#1}{$\vcenter{\hrule height 1.2pt width 8mm}$}}
\usepackage{graphicx,mwe}

\usepackage{footmisc}
\usepackage{epsf,slashed,soul}
\usepackage{pifont}
\usepackage{bbold}
\usepackage{dsfont}
\usepackage{tikz}
\usepackage{verbatim}
\usepackage{placeins}
\hypersetup{
   colorlinks,
   linkcolor={red!50!black},
    citecolor={blue!50!black},
    urlcolor={blue!80!black}
}

\DeclareSymbolFont{usualmathcal}{OMS}{cmsy}{m}{n}
\DeclareSymbolFontAlphabet{\mathcal}{usualmathcal}

\newcommand{\bea}{\begin{eqnarray}}
\newcommand{\eea}{\end{eqnarray}}

\newcommand{\abs}[1]{|#1|}

\newcommand{\be}{\begin{equation}}
\newcommand{\ee}{\end{equation}}

\newcommand{\cd}{c^\dagger}
\newcommand{\su}{\mathfrak{su}(1,1)}

\title{Krylov meets Fock: Complementary lenses on quantum dynamics}

\author[a,b]{Giuseppe Di Giulio,}
\author[c]{Dimitrios Patramanis}

\affiliation[a]{Physikalisches Institut, Universität Bonn, Nußallee 12, 53115 Bonn, Germany}
\affiliation[b]{The Oskar Klein Centre and Department of Physics, Stockholm University, AlbaNova, 106 91 Stockholm, Sweden}
\affiliation[c]{Kavli Institute for Theoretical Sciences,
  University of Chinese Academy of Sciences, Beijing 100190, China}

\emailAdd{giuseppe.di-giulio@fysik.su.se}
\emailAdd{patramanisdimitrios@gmail.com}

\abstract{Quantum dynamics in Krylov space can be represented by a particle hopping on a chain whose sites are the Krylov basis
vectors. The goal of this work is to extend this idea to the Fock basis, which
in turn allows the description of a generic dynamics in terms of a more general, possibly higher-dimensional lattice. This construction has been dubbed the Fock state lattice (FSL), since its sites are identified with Fock states. To analyze the dynamics on the FSL we introduce various measures of the spread of a state, comparing them among themselves and with their Krylov counterparts. We apply our analysis to two models of light-matter interaction
studied in quantum optics: the multi-mode Jaynes-Cummings and the Dicke models.
We find that, whenever the FSL is not a chain, the corresponding dynamics
provides insights complementary to the Krylov space analysis and, in some
instances, greater analytical control. For example, we obtain analytical results
for the FSL spread measures of the Dicke model in the thermodynamic limit and determine that
their scaling in time detects the superradiant phase transition. }

\begin{document}

\maketitle

\section{Introduction}

Characterizing the dynamics of quantum systems remains to this day an active and vibrant research topic despite a whole century separating us from the nascence of quantum theory. This period has undoubtedly been rife with important advances, yet questions regarding quantum chaos, thermalization, and phase transitions persist, especially when they arise in tandem with problems at the frontier such as strongly coupled systems, black holes and more generally complex systems.

Krylov space methods, recently revived in this context, make it possible to
represent a generic unitary quantum dynamics as the evolution of a particle
hopping on a one-dimensional chain. This representation is particularly useful
for visualizing and quantifying important features of quantum evolution, such
as the complexity and spread within the Hilbert space of a system. While
these methods are efficient to be implemented numerically, they are not always
analytically tractable, in particular for systems with high-dimensional Hilbert
spaces. This motivates the search for alternative, and possibly complementary,
ways of representing quantum dynamics on graphs and lattices, together with the
tools needed to study them. In this work, we analyze the Fock state lattice
(FSL) approach from this point of view. We study how to quantify the spread of
the evolving state and its distance from the initial one, comparing
the resulting picture with the one obtained through the Krylov
space approach.
Naturally, the two frameworks share the same logic. A quantum evolution is recast
as the motion of a particle on a lattice whose sites are the elements of a
basis, and the two differ only in which basis is chosen, i.e. the Fock basis for
the FSL, the Krylov basis for the Krylov chain. It is therefore natural to ask how the two resulting lattices differ, and what
each of them reveals about the same quantum evolution.
As our testbed we  will use a class of quantum optics models that concern the interaction of atoms (qubits) and radiation. These play an important role both as toy models for high-energy and condensed-matter applications, and as experimental platforms \cite{Kirton:2019jes,Larson:2021flt, Forn-Diaz:2019epp, FriskKockum:2018nuz,Gerritsma:2009doz,Lewis-Swan:2018sdr}.

While we will review both the Krylov and FSL frameworks in the following section, it is important to highlight their utility and several of their applications.
The use of Krylov space methods with the purpose of probing the dynamics of quantum systems was popularized by the authors of \cite{Parker:2018yvk}. Many of the relevant computational techniques they based their seminal work on can be found for example in \cite{viswanath1994recursion}. In recent years there has been an explosion of activity on the topic with works ranging from condensed matter \cite{Rabinovici:2021qqt,Noh:2021vnc,Bhattacharjee:2022qjw,Caputa:2025ucl,Capizzi:2025eyg, Jiang:2026ejp} to high energy applications \cite{Dymarsky:2021bjq,Avdoshkin:2022xuw,Caputa:2021ori,Camargo:2022rnt,Rabinovici:2023yex,Caputa:2023vyr,Ambrosini:2024sre,Caputa:2024sux,Heller:2024ldz,Caputa:2025dep,Heller:2025ddj,Aguilar-Gutierrez:2025hty,Ambrosini:2025hvo,Kristjansen:2025xqo,Bhattacharya:2026ybq,Graef:2026uaz}, also generating a significant discourse relating to quantum chaos \cite{Rabinovici:2020ryf,Rabinovici:2022beu,Bhattacharjee:2022vlt,Balasubramanian:2022dnj,Bhattacharyya:2023grv,Camargo:2023eev,Balasubramanian:2023kwd,Camargo:2024rrj,Camargo:2024deu,Chen:2024imd,Beetar:2025erl,Basu:2025ubf,Bhattacharjee:2024yxj,Angelinos:2025drf,Das:2024tnw,Murugan:2026rfa,Aguilar-Gutierrez:2026jjv,Erdmenger:2026iga}. At the time of writing there are already a few review articles that one can refer to for the most important advances and the state of the art \cite{Baiguera:2025dkc, Nandy:2024evd, Rabinovici:2025otw,Jeong:2026gdc}. 
A line of works that is of exceptional importance for our present purposes is related to the universal aspects that a Krylov space inherits from the symmetry structure of a quantum system \cite{Caputa:2021sib,Patramanis:2021lkx,  Balasubramanian:2022tpr, Hornedal:2022pkc, Caputa:2022zsr, Haque:2022ncl, Patramanis:2023cwz, Caputa:2025mii,Chowdhury:2024qaj, Caputa:2025ozd,Balasubramanian:2025xkj,Grabarits:2026hjz} and the efficacy of Krylov-related quantities in detecting quantum phase transitions \cite{Caputa:2022eye,Caputa:2022yju,Afrasiar:2022efk,Bento:2023bjn,Baggioli:2024wbz,Grabarits:2025xys}.

While FSLs have not garnered the same level of attention that Krylov-space-related methods have, they have nonetheless proven to be important concepts, especially as potential platforms for the simulation of condensed matter systems with interesting topological properties \cite{Cai:2020vfx,Zhao:2024xxq,Naves:2025iec}. This line of research has already found certain experimental realizations \cite{Deng:2022ilg,Zhang:2024auk}, which makes it an appealing direction as a potential bridge between theory and experiment. Here we are interested in the use of this framework for the formal representation of quantum dynamics on a lattice and as such we will mainly follow the paradigm set by \cite{Saugmann:2022qzh,Ferraro:2026twt}, in which the authors describe in great detail the construction of FSLs for a variety of models and discuss aspects of their dynamics.

In our work, we will focus on quantum optics models and in particular the Jaynes-Cummings (JC) and Dicke models in their various regimes. In brief, the former concerns the interaction of a single atom with multiple modes of the electromagnetic field whereas the latter reverses the problem and describes the interaction of a collection of atoms with a single mode. Both have been studied extensively for a variety of experimental applications including cavity and circuit QED, and trapped ions (see for example \cite{Larson:2021flt,Kirton:2019jes} and references therein). More importantly for our purposes, however, these models and their close relatives (Rabi, Tavis-Cummings) constitute the ideal theoretical playground for the study of quantum dynamics, where the interplay between notions like integrability, chaos, and thermalization unfolds in a controlled setting. Their importance in this regard cannot be overstated, given that they have served in this capacity for decades producing a host of related works (see for example the highly influential \cite{Eberly:1980zz,GeaBanacloche1990CollapseAR,Braak:2011tmj,emary2003chaos}).

Having laid out all the ingredients we will be using henceforth, we have chosen to organize the rest of this work as follows. In Sec.\,\ref{review section} we provide a short review of the Krylov and FSL frameworks, including all the necessary definitions and concepts that will appear in later sections. In Sec.\,\ref{intro to qom} we introduce the quantum optics models we will be focusing on and use the simple examples of the Rabi and single-mode JC models as a springboard for the more involved analysis we perform in Secs.\,\ref{multi-mode} and \ref{Dicke}, which concern the multi-mode JC and Dicke models respectively. Finally, we provide a summary of our conclusions along with potential future directions in Sec.\,\ref{conclusions}.

\section{The Krylov and FSL frameworks} \label{review section}

In this section we provide brief reviews of the topics of FSLs and Krylov complexity in order to cover the basic notions that will appear repeatedly throughout the rest of this work. Even though the two frameworks first appeared around the same time and they are, as we will discuss in later sections, intimately related, there has been little overlap between them since, until now, they have been used toward different ends.

\subsection{Krylov space and complexity}
\label{subsec:Krylov}

The simple yet powerful idea that underlies this framework is that one can reduce the problem of tracking operator growth or quantum state evolution to a one-dimensional problem. This is achieved by tracking the system evolution in the so-called Krylov basis rather than the full Hilbert space. The former is defined through the repeated application of the Hamiltonian to an initial state, followed by the implementation of an orthonormalization procedure. In other words starting with the time evolution of a state
\be
e^{-iHt}\ket{\psi_0}= \sum_{k=0}^{\infty}\frac{(-i t)^k}{k!}H^k \ket{\psi_0}~,
\ee
we can create a basis from the vectors
\be
\{ \ket{\psi_0}, H \ket{\psi_0}, H^2 \ket{\psi_0},...\} ~,
\ee
which we can then use to track the evolution of the system. Of course, this requires that these vectors are orthonormalized. This is typically done through a Gram-Schmidt process that, for spaces generated by repeated application of an operator to a seed vector, is also known as the Lanczos algorithm.
What makes this procedure special is that the Hamiltonian, expressed in the resulting orthonormal basis, is tridiagonal.
More concretely, the recursive construction of the Krylov basis follows the steps below.
\begin{enumerate}
    \item The initial state $\ket{\psi_0}$ is identified with the first Krylov basis vector which we denote $\ket{K_0}$. This is simply the initial condition of the problem.
    \item We apply the Hamiltonian to the initial state obtaining 
    \be
    H\ket{K_0}=\ket{A_1} +a_0 \ket{K_0}~,
    \ee
    where $a_0=\bra{K_0}H\ket{K_0}$.
    We then define the second Krylov basis vector as $\ket{K_1}= b_1^{-1}\ket{A_1}$,
where  $b_1 =\sqrt{\braket{A_1|A_1}}$ is the normalization constant. 
\item We repeat this process with each subsequent Krylov state.
The action of $H$ on the basis vector $|K_{k}\rangle$ takes the form
\be
H \ket{K_k}=a_k\ket{K_k} +b_k\ket{K_{k-1}} + b_{k+1}\ket{K_{k+1}}~,
\ee     from which $|K_{k+1}\rangle$ is recursively obtained.
\end{enumerate}
After orthonormalization one is left with the Krylov basis vectors $\{\ket{K_k}\}$ and the matrix elements of the Hamiltonian $a_k$ and  $b_k$, conventionally denoting the diagonal and off-diagonal elements respectively and in this context called the {\it Lanczos coefficients}. 

It is now possible to express the time evolved state as
\be
\ket{\psi(t)}=\sum_{k=0}^{\mathcal{K}-1} \varphi_k(t) \ket{K_k}~,
\ee
where $\mathcal{K}$ is the cardinality of the Krylov basis.
The evolution of the coefficients $\varphi_k$ is what makes transparent the reduction in the dimensionality of the problem, since it can be easily shown to be 
\be
\label{eq:evolutionKamplitude}
i\frac{\partial \varphi_k(t)}{\partial t}= a_k\varphi_k(t)+b_k\varphi_{k-1}(t)+b_{k+1}\varphi_{k+1}(t)~.
\ee
Notice that this is a (discrete) Schr{\"o}dinger equation describing a quantum particle on a finite or semi-infinite chain. This is commonly referred to as the \textit{Krylov chain} and it provides an insightful picture into the dynamical processes of the quantum system under consideration. Krylov complexity is defined as the average position of the particle on said chain
\be
\label{eq:def_spreadcomplexity}
C_K=\sum_{k=0}^{\mathcal{K}-1} k \abs{\varphi_k}^2~,
\ee
which essentially quantifies the average power of the Hamiltonian one needs to apply to the initial state in order to effectively describe the time evolved state. 
When it refers to the evolution of states, as in the case we are reviewing,
this quantity is often called spread complexity. Here, we prefer the name Krylov
complexity, to stress the key role played by the Krylov basis in this
construction.

Once the probabilities $\abs{\varphi_k}^2$ associated with the Krylov basis are
known, it is very natural to compute the Shannon entropy of the resulting
distribution, usually referred to in this context as the K-entropy \cite{Barbon:2019wsy}
\begin{equation}
\label{eq:Kentropy}
    S_K=-\sum_{k=0}^{\mathcal{K}-1} \abs{\varphi_k}^2\ln \abs{\varphi_k}^2\,.
\end{equation}
This quantity measures the spread of the evolving state over the Krylov basis.
In the next sections, we compare $S_K$ with an analogous notion of entropy
describing the spread in the Fock basis.

\subsection{Fock state lattices}

In the previous section we saw that the dynamics of any quantum system can in principle be treated using the Krylov methods we reviewed. However, experience teaches us that this is hardly an analytically tractable problem outside a limited class of systems with exceptional symmetry properties. Put simply, reducing the dynamics of a system to a one-dimensional problem is a marvelous technique when it can be brought to bear, but in reality one often has to resort to numerics or approximations. For this reason we wish to put forth FSLs as a closely related alternative that naturally occurs in a variety of settings where the Krylov framework is of limited utility. 

The term Fock state lattice was coined in \cite{Wang:2016ciq} in relation to the Jaynes-Cummings model, though a more general picture regarding the utility of this approach to different quantum optics models was provided in \cite{Saugmann:2022qzh}, which we will closely follow in introducing the relevant concepts. Recently a more abstract approach regarding the algebraic properties of FSLs and their construction was given in \cite{Ferraro:2026twt}, which will also be of use in understanding the relation to Krylov complexity. 

Let us, however, start with the fundamental question. What is a Fock state lattice? On a heuristic level, it constitutes a representation of the Fock space of a given system where every Fock state is represented by a vertex. The Hamiltonian of the system describes the allowed transitions between Fock states, which in turn manifest as edges between the vertices. In this way we are left with a lattice structure that describes the connectivity between different Fock states and can be used as an effective tool to probe the dynamics of a quantum system under Hamiltonian evolution. 
In this manuscript, we use the term Fock states for the elements of any basis
labeled by integers, such that one can move from one element to another by
applying raising and lowering operators. This family obviously includes the
energy eigenstates of the harmonic oscillator, but also the eigenstates of
$\sigma^{(z)}$, $\ket{\uparrow}$ and $\ket{\downarrow}$.
In the quantum-optics models we are interested in, the spin-$1/2$ states
describe two-level atoms, while the harmonic oscillators represent the modes of the photons
interacting with them.
On a more general note, these states constitute a unitary irreducible representation of an underlying symmetry algebra. This idea has been expanded upon in \cite{Ferraro:2026rrq}, and, while this perspective remains mostly beyond our scope, we make use of this fact in Appendix \ref{Krylov-Fock equivalence}.

\begin{figure}[ht]
    \centering
\includegraphics[width=0.75\linewidth]{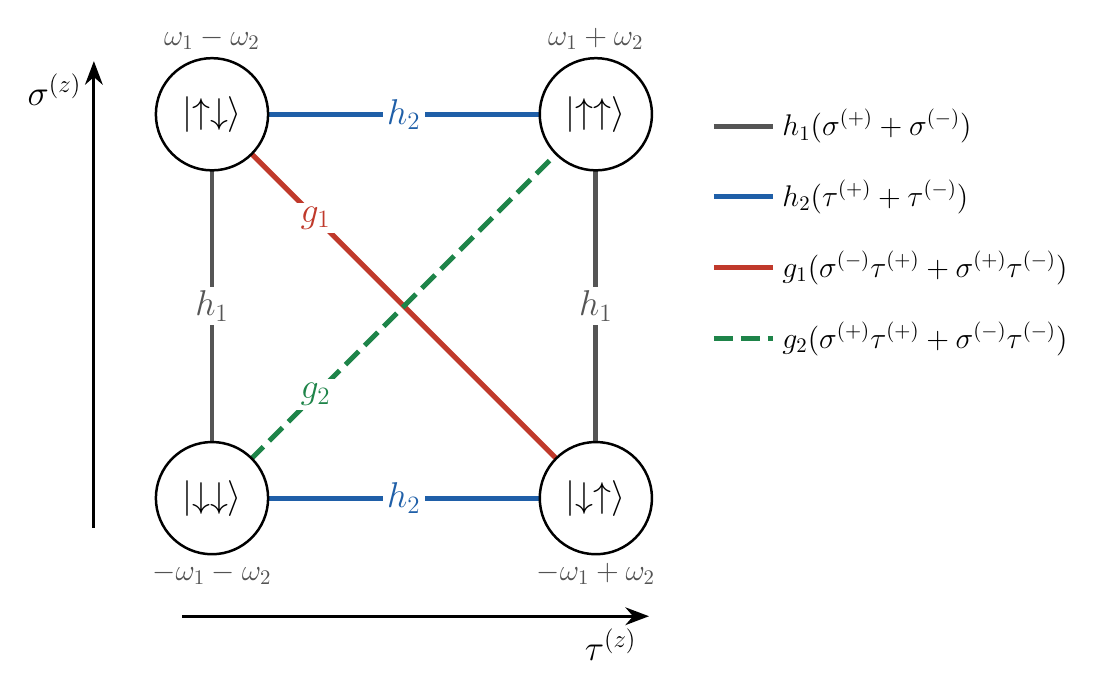}
    \caption{The FSL characterizing the interaction  of two spin-1/2 degrees of freedom given by the Hamiltonian \eqref{eq:Example_Ham}. It contains four vertices (Fock states), all of which are connected to each other.}
    \label{simplest FSL}
\end{figure}

To become familiar with the notion of FSL and with its construction, we consider
as a first example the interaction between two spin-$1/2$ degrees of freedom
represented by the Pauli matrices $\sigma^{(\alpha)}$ and $\tau^{(\alpha)}$, with $\alpha=x,y,z$, described by the Hamiltonian
\begin{eqnarray}
\label{eq:Example_Ham}
H_{FSL}&=& \omega_1 \sigma^{(z)}+\omega_2 \tau^{(z)} 
+
h_1 \left(\sigma^{(+)}+\sigma^{(-)}\right)+h_2 \left(\tau^{(+)}+\tau^{(-)}\right)
\\
&+&g_1 \left( \sigma^{(-)}\tau^{(+)} +\sigma^{(+)}\tau^{(-)}\right)+ g_2 \left( \sigma^{(+)}\tau^{(+)} +\sigma^{(-)}\tau^{(-)}\right)~,
\nonumber
\end{eqnarray}
where $\sigma^{(\pm)}=\tfrac{1}{2}\big(\sigma^{(x)}\pm i\,\sigma^{(y)}\big)$ and
$\tau^{(\pm)}=\tfrac{1}{2}\big(\tau^{(x)}\pm i\,\tau^{(y)}\big)$ are the ladder operators of the spin variables. The corresponding FSL is illustrated in Fig.\,\ref{simplest FSL}. To construct it, we draw an edge between any two Fock states connected by a
non-vanishing off-diagonal matrix element of the Hamiltonian, whose value fixes
the weight of that edge. In this example, the  four Fock states are $\ket{\downarrow,\downarrow},\ket{\downarrow,\uparrow}, \ket{\uparrow,\downarrow},\ket{\uparrow,\uparrow}$, each corresponding to a combination of the possible states of the spins.

This simple but instructive example already allows us to appreciate the
prominent role that symmetries play in the construction of the FSL. Indeed, the Hamiltonian \eqref{eq:Example_Ham} has no symmetry.
To endow the model with one, we switch off some of its couplings and discuss
the consequences on the FSL of Fig.\,\ref{simplest FSL}. By setting
$h_1=h_2=0$, we introduce the $\mathbb{Z}_2$ symmetry. In terms of
the FSL, this reduces the highly connected graph of Fig.\,\ref{simplest FSL} to
a graph with two disconnected components, each made of only two states. In the
representation of Fig.\,\ref{simplest FSL}, the two disconnected components are the two diagonal red and dashed green edges. Each component corresponds to one of the two
$\mathbb{Z}_2$ parity sectors.

As we will see in other examples below, adding
symmetries generally decomposes the FSL into smaller disconnected components,
associated with the charge sectors of the symmetry. From the point of view of the dynamics on the FSL
this is a simplification: as expected in the presence of a symmetry, an evolution
starting on a given component remains in that component, and therefore explores
a smaller portion of the full Hilbert space.

It should be fairly obvious that this example does not provide a very realistic model of the interaction between matter and radiation. In the sections to follow we will see the FSLs that more realistic Hamiltonians produce, though their construction relies on the same principles as this elementary setup.

\subsection{A preliminary comparison}

In later sections we will investigate the differences between the Krylov and FSL frameworks in depth through a series of examples. However, there are a few rudimentary differences that we believe are worth emphasizing before we set out on this path. Perhaps the most striking difference has to do with the sensitivity to initial conditions or lack thereof. It is well known that the construction of the Krylov basis is highly dependent on the choice of initial state. Depending on the situation this can be interpreted as a feature or an obstacle, though the fact remains that different initial states of the same system can potentially lead to completely different Krylov spaces and therefore complexities. 

On the other hand, the FSL is a fixed property of the system given its Hamiltonian. The initial state is then one vertex (or superposition of vertices) on that predetermined structure. Unlike the Krylov chain where the simple notion of average distance serves as a powerful diagnostic tool, the FSL is characterized by an inherent ambiguity when it comes to defining a notion of spread or distance. This is a generic feature of lattices that are anything other than a chain or loop, and as such this is an ambiguity we will have to contend with. In addition, the FSL as a graph, including its connectivity, is strongly model-dependent. Thus, any comparison between results from the FSL approach applied to different models can be subtle.

To conclude this brief comparison, we stress that for some examples the FSL and Krylov approaches provide the same description. This becomes transparent in the case of a system with dynamical symmetry given by a Lie algebra with one set of ladder operators. This includes symmetries such as $\mathfrak{su}(1,1),\, \mathfrak{su}(2)$ and Heisenberg-Weyl which have been studied extensively in this context. For these cases and assuming the initial state is a highest weight state of the representation, the FSL and Krylov chain are identified. In other words the Krylov space generated from such a state is precisely the Fock space. This idea has been implicitly used in a significant portion of the Krylov literature (see, e.g, \cite{Caputa:2021sib,Patramanis:2021lkx,Hornedal:2022pkc,Grabarits:2026hjz}) though to our knowledge it has not been clearly stated before. As such we take this opportunity to make it explicit and provide the accompanying proof in Appendix \ref{Krylov-Fock equivalence}.

\section{Simple models of light-matter interaction} \label{intro to qom}

In this section we review two models commonly studied in quantum optics, which will serve as the testing ground for the rest of the manuscript.
    We begin with the simplest setup describing the interaction between a two-level atom and a single mode of the electromagnetic field, which is known in the quantum optics community as the quantum Rabi model \cite{PhysRev.51.652}. The relevant  Hamiltonian is given by 
    \be
    H_{\text{Rabi}}=\omega a^\dagger a +\frac{\Delta}{2} \sigma^{(z)} +g\sigma^{(x)} \left(a^\dagger +a\right)~.
    \ee
While it looks simplistic, this model has a very rich structure that over the years has led to important insights regarding both theory and experiment (for a review see \cite{Lee:2017dwn}). It is convenient for our purposes to rewrite the Hamiltonian with the final term expanded as
\be \label{Rabi Hamiltonian}
  H_{\text{Rabi}}=\omega a^\dagger a +\frac{\Delta}{2} \sigma^{(z)} +g \left(\sigma^{(+)} a +\sigma^{(-)}a^\dagger \right) +g \left(\sigma^{(+)} a^\dagger +\sigma^{(-)} a \right)~.
\ee
The penultimate term is referred to as ``rotating" and the last one as ``counter-rotating". Before investigating the dynamics of this model itself, it is worth noting the different ways in which it can be generalized since these will serve as the topics of later sections. 

If instead of a single atom in the Rabi Hamiltonian we consider an arbitrary number of them, we obtain the Dicke model \cite{Dicke_or} with Hamiltonian 
\be \label{Dicke Hamiltonian}
H_\text{Dicke}=\omega a^\dagger a + \frac{\Delta}{2} \sum_{j=1}^M\sigma_j^{(z)} +\frac{ g }{\sqrt{M}}\left(a+a^\dagger\right)\sum_{j=1}^{M} \left(\sigma_j^{(+)} +\sigma_j^{(-)}\right)~,
\ee
where $M$ denotes the number of atoms. 

If instead we consider the so-called rotating wave approximation (RWA) where one discards the counter-rotating term, what we are left with is the Jaynes-Cummings (JC) model  \cite{Jaynes:1963zz} with Hamiltonian
\be
\label{eq:JCham_1mode}
H_\text{JC}=\omega a^\dagger a +\frac{\Delta}{2} \sigma^{(z)} +g \left(\sigma^{(+)} a +\sigma^{(-)}a^\dagger \right) ~.
\ee
This can be further extended by considering multiple modes of the electromagnetic field, leading to the multi-mode JC model 
\be 
\label{eq:JCham_Nmode}
H^{(N)}_\text{JC}=\frac{\Delta}{2} \sigma^{(z)} + \sum_{i=1}^N\omega_i a_i^\dagger a_i +\sum_{i=1}^N g_i\left(\sigma^{(+)} a_i +\sigma^{(-)}a_i^\dagger\right)~.
\ee

Even though there are many names involved, this class of models simply describes the interaction of light and atoms in various regimes. We hope that this short exposition clarifies what situation each model is attempting to capture. 

\begin{figure}[ht]
    \centering
    \includegraphics[width=0.75\linewidth]{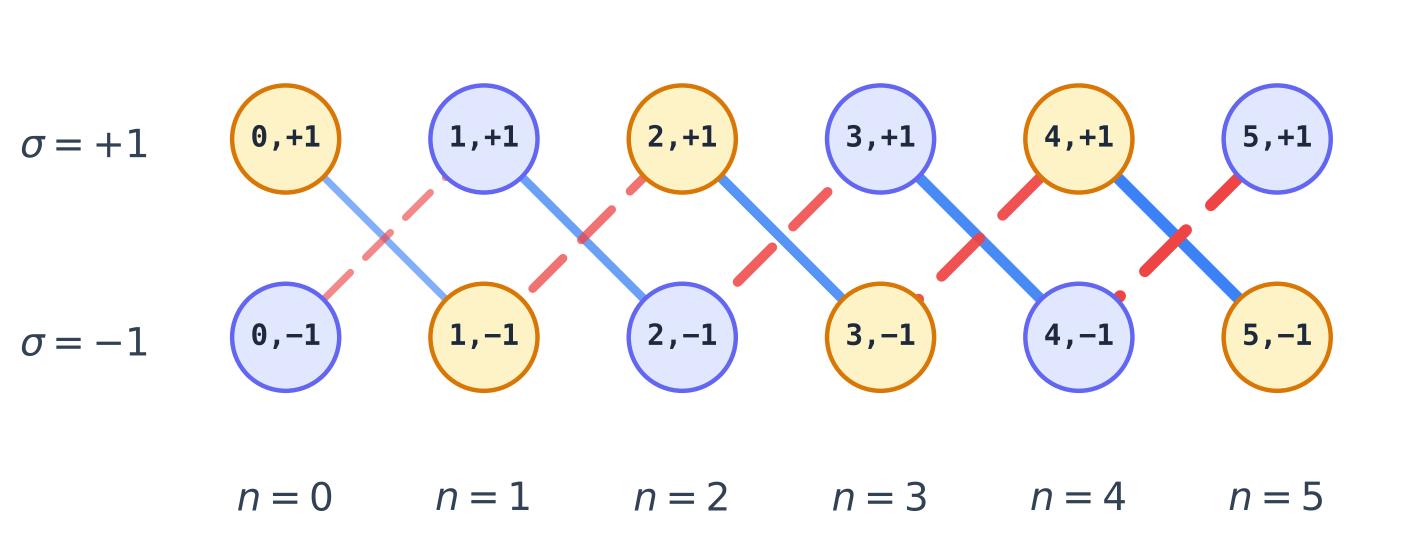}
    \caption{The FSL of the quantum Rabi model. 
    The vertices in the upper row correspond to states with excited atomic level (spin up), while the ones in the bottom row are in the atomic ground state (spin down).
    The blue edges correspond to the rotating and the red, dashed edges correspond to the counter-rotating terms of the Hamiltonian \eqref{Rabi Hamiltonian}. Note that the thickness of both kinds of edges increases as $n$ grows, signifying the increase in the associated hopping amplitudes.}
    \label{Rabi fsl}
\end{figure}

For the rest of this section we will focus on the quantum Rabi and single-mode JC models. The former's FSL structure has been described in \cite{Saugmann:2022qzh}, where the authors determined it to be composed of two disconnected chains, each associated with a parity subsector of the Hamiltonian \eqref{Rabi Hamiltonian} as seen in Fig.\,\ref{Rabi fsl}.

When it comes to the Krylov space of this system, it is easy to see that if we choose as an initial state the vacuum of the photonic mode times either the ground ($\sigma=-1$) or excited ($\sigma=1$) state of the atom, then the Hamiltonian \eqref{Rabi Hamiltonian} is automatically in tridiagonal form and the Krylov chain of the resulting dynamics is given by one of the two FSL components in Fig.\,\ref{Rabi fsl}, depending on whether the initial state has $\sigma=-1$ or $\sigma=1$. The Lanczos coefficients follow immediately as 
\be \label{eq:Rabi Lanczos}
a_k= \omega k + (-1)^k\sigma \frac{\Delta}{2}, \quad b_k=g\sqrt{k}~.
\ee
This type of growth for $b_k$ as a function of $k$, in particular, is associated with integrable dynamics. 
Even though this set of Lanczos coefficients does not fit into a known universality class (because of the staggering of $a_k$) and a closed expression for the complexity is missing, we can get some insight into the inner workings of the Rabi model by considering the  $\Delta=0$ limit, in which the Krylov complexity takes the form dictated by Heisenberg-Weyl symmetry \cite{Balasubramanian:2022tpr}
\be
\label{eq:complexity Rabi}
C_K(t)=\frac{4g^2}{\omega^2}\sin^2{\left(\frac{\omega t}{2}\right)}~.
\ee
In the limit where $\omega$ is small, i.e. in the semi-classical limit, we recover the characteristic quadratic growth associated with this class of Lanczos coefficients
\be
C_K(t)=g^2t^2~.
\ee

The results \eqref{eq:Rabi Lanczos} and \eqref{eq:complexity Rabi} appear consistent with the Rabi model being integrable \cite{Braak:2011tmj,Larson:2013isl}. In other words, despite its Hilbert space being infinite dimensional, the Rabi model shows no signs of fast scrambling or chaos, at least as far as the Krylov and FSL quantities are concerned. The simplicity of this example makes it such that the FSL and Krylov chains are identified, assuming the initial state is one of the ends of the chains in Fig.\,\ref{Rabi fsl}. However, it should be noted that if the initial seed is a superposition of states in the two subsectors, the reduction to the Krylov basis is not as straightforward.

Let us move on to the single-mode JC model described by the Hamiltonian \eqref{eq:JCham_1mode}. Due to the U(1) symmetry generated by \begin{equation}
\label{eq:JCmodeQ}
    Q_1=\frac{\sigma^{(z)}+1}{2}+a^\dagger a\,,
\end{equation} 
the Hilbert space of this model decomposes into infinitely many charge sectors labeled by the eigenvalues of $Q_1$, 
which are the non-negative integers $q_1=n+(\sigma+1)/2$, where $n$ is the bosonic occupation number, i.e. the photon number, and $\sigma=\pm 1$ gives the atomic excitation.
Each sector contains only two states given that, to keep $q_1$ constant, we either increase it by removing a photon, or decrease it by adding one. This leads to infinitely many disconnected two-site FSLs.

\begin{figure}[ht]
    \centering
    \includegraphics[width=0.99\linewidth]{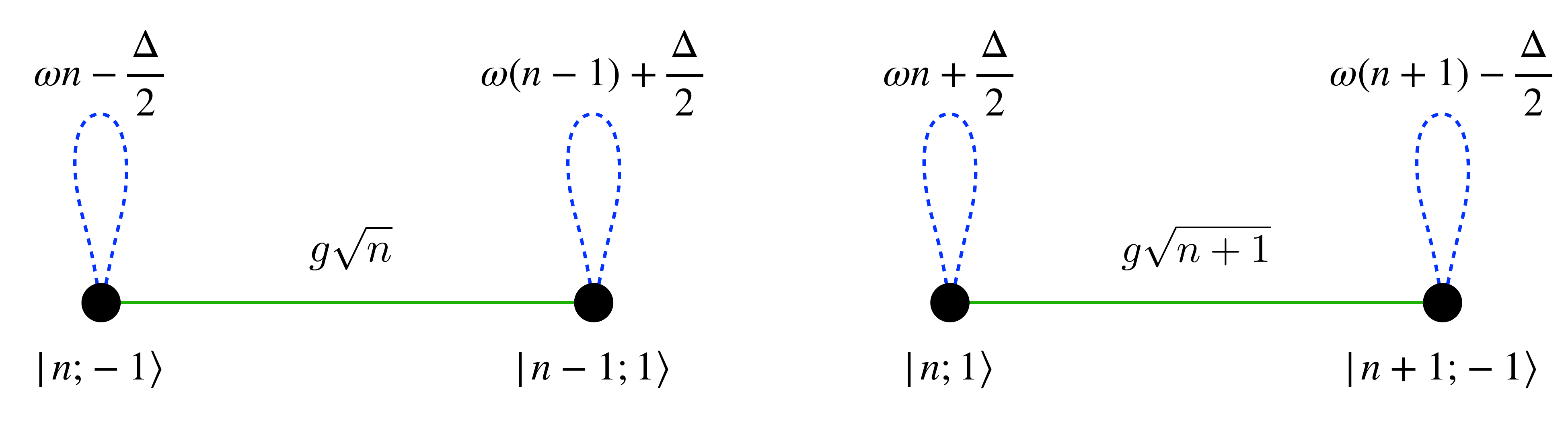}
    \caption{Fock state lattice components of the single-mode JC model with Hamiltonian \eqref{eq:JCham_1mode}. Different two-site lattices correspond to different sectors with respect to the charge \eqref{eq:JCmodeQ}. The ones represented correspond to the sectors with $q_1=n$ (left panel) and $q_1=n+1$ (right panel).}
    \label{fig:JC_1mode_FSL}
\end{figure}

Since each FSL is made of only two sites, the structure is simple and can be seen as a two-site chain, where coupling and on-site energy are given by the action of the Hamiltonian on the Fock states (see Fig.\,\ref{fig:JC_1mode_FSL}). Also in this instance, the FSL has a direct connection with the Krylov chain. Indeed, by considering the dynamics induced by \eqref{eq:JCham_1mode} on a certain initial Fock state $|n;\sigma\rangle$, the arising Krylov chain where this evolution takes place is the FSL associated with the charge sector to which the initial state belongs.
In this case, it is also possible to solve the dynamics on the Krylov basis, i.e. the Fock basis, and compute the Krylov complexity.
Following the procedure described in Sec.\,\ref{subsec:Krylov}, we can compute the Lanczos coefficients of the dynamics, which read
\begin{equation}
\label{eq:Lanczos1modeJC_1}
    a_0=\omega n+\sigma\frac{\Delta}{2}\,,\qquad
    b_1=g\sqrt{q_1}\,,
    \qquad
     a_1=\omega (n+\sigma)-\sigma\frac{\Delta}{2}\,.
\end{equation}
Solving the hopping model evolution and extracting the Krylov complexity, we obtain
\begin{equation}
    C_K(t)=\frac{g^{2}q_1}{\nu^{2}+g^{2}q_1}\,
    \sin^{2}\!\left(\sqrt{\nu^{2}+g^{2}q_1}\;t\right),
    \qquad
    q_1=n+\frac{1+\sigma}{2}\,,\qquad \nu\equiv\frac{\Delta-\omega}{2}\,.
\end{equation}
The identification of the Krylov chain with the FSL in a given charge sector is valid only if we focus on the dynamics starting from a Fock state. Indeed, if the initial state is a linear combination of the two Fock states in the same charge sector, it is straightforward to verify that also the second Krylov basis vector is a linear combination of them. As a result, the Krylov chain is a two-site chain, but the states represented by the vertices are different from those in the FSL. 

This difference is even more evident if the initial state is a linear combination across different sectors. In that case, the Krylov chain can contain (at most) as many vectors as the sum of cardinalities of the charge sectors involved in the initial state.

The simple examples discussed in this section already show a central point of this work: The FSL and the Krylov chain of a given model are generally different lattice structures that we can use to describe dynamical processes and they coincide only in specific instances. As such, they contain different physical information. One of the goals of this paper is to identify it and understand how to exploit the two approaches in different circumstances.

\section{Multi-mode JC model} \label{multi-mode}

In this section, we consider the multi-mode JC model described by the
Hamiltonian \eqref{eq:JCham_Nmode}. We study it analytically and numerically
for $N=2$ and $N=3$, from both the Krylov space and the FSL perspectives,
comparing the two descriptions and the results from the two approaches.

\subsection{The model and its Fock state lattice}

Let us begin by reviewing the construction of the FSL of the JC model with Hamiltonian \eqref{eq:JCham_Nmode}. Generalizing the discussion in Sec.\,\ref{intro to qom}, the Fock basis that we use to study the dynamics of our system through the FSL contains vectors of the form $\ket{n_1,n_2, \dots,n_N;\sigma} $, with $n_i$ non-negative integers and $\sigma= \pm 1$.

The first step to understand the FSL of the model is the analysis of its symmetries. It is straightforward to verify that the Hamiltonian \eqref{eq:JCham_Nmode} commutes with the operator
\begin{equation}
\label{eq:JCmodeQ_Nmode}
    Q_N=\frac{\sigma^{(z)}+1}{2}+\sum_{i=1}^Na_i^\dagger a_i\,,
\end{equation}
which counts the occupation number in all the bosonic modes, i.e. the total number of photons, plus the atomic excitation. This operator and its eigenvalues $q_N$ generalize $Q_1$ in \eqref{eq:JCmodeQ} and $q_1$ to the multi-mode model.  In this case, the atomic excitation and the number of photons are not conserved separately due to the last two terms in \eqref{eq:JCham_Nmode}, which create or destroy a photon and at the same time flip the spin variable. Since the eigenvalues of $Q_N$ are non-negative integer numbers, the FSL of the JC model is made of countably infinitely many disconnected sublattices, each labeled by a charge eigenvalue $q_N=0,1,2,\dots$. As will become evident in the next subsections, the structure of the disconnected components of the FSL depends on $N$ and becomes richer as the number of bosonic modes increases.

The {\it large-detuning limit} is the regime where the atomic gap $\Delta$ sets the fastest time scale compared with those associated with all the spin-photon couplings $g_i$ and the mode frequencies, i.e. $\vert\Delta\vert\gg g_i\sqrt{q_N},\, \omega_i$, where $q_N$ is the charge of the symmetry sector to which we restrict our analysis.
Under this assumption, the dynamics is effectively governed by the photons,  leading to the perturbative picture describing the effect of the spin variable as mediating virtual absorptions and emissions of photons. Formally, using adiabatic elimination \cite{Zoller:1997su}, we obtain the effective Hamiltonian
\begin{equation}
    H_{\rm det}=\frac{\Delta}{2}\,\sigma^{(z)}
+\sum_{i=1}^{N}\omega_i\,a^\dagger_i a_i
+\frac{\sigma^{(z)}}{\Delta}\sum_{i,j=1}^{N}g_i g_j\,a^\dagger_i a_j
+\frac{1+\sigma^{(z)}}{2}\,\frac{\sum_{i=1}^{N}g_i^2}{\Delta}\,,
\label{eq:multimodeJC_largedet}
\end{equation}
where the spin and the photonic sectors are now decoupled as expected. 

Interestingly, the symmetries of the JC model change in this regime. Indeed, the number of photons and the atomic excitation are now separately conserved. This implies that the disconnected sublattices of the FSL are, again, countably infinitely many, as they are labeled by the number of photons $\mathcal{N}_N$, but each of them appears twice due to the two values of the atomic excitation labeled by $\sigma=\pm 1$. As we discuss in the next subsections, the structure of FSL sublattices at fixed $\mathcal{N}_N$ is different from that of the FSL sublattices at fixed $q_N$ away from the large-detuning regime. In other words, apart from doubling the number of FSL components, the large-detuning limit also modifies the structure of these sublattices.

\subsection{Two-mode JC model}
\label{subsec:twomode_JC}

 The first multi-mode JC model is defined by the Hamiltonian \eqref{eq:JCham_Nmode} with $N=2$. We begin our analysis by quickly reviewing the construction of the FSL.
 
 As explained in the previous section, the FSL of this model has infinitely many disconnected components, each labeled by the sum of atomic excitation and photon number, i.e. $q_2=\frac{\sigma+1}{2}+n_1+n_2=0,1,2\dots$.
Let us consider one of these charge sectors and let the Hamiltonian of the model act on a generic vector in that sector. We obtain
\begin{eqnarray}
\label{eq:H2JConFock_minus}
    H^{(2)}_\text{JC}\vert n_1,n_2;-1 \rangle&=&\left(-\frac{\Delta}{2}+\omega_1n_1+\omega_2n_2\right)\vert n_1,n_2;-1 \rangle+g_1\sqrt{n_1}\vert n_1-1,n_2;1 \rangle
    \\
    \nonumber
    &+&
g_2\sqrt{n_2}\vert n_1,n_2-1;1 \rangle\,,
    \\
    H^{(2)}_\text{JC}\vert n_1,n_2;1 \rangle&=&\left(\frac{\Delta}{2}+\omega_1n_1+\omega_2n_2\right)\vert n_1,n_2;1 \rangle+g_1\sqrt{n_1+1}\vert n_1+1,n_2;-1 \rangle
    \label{eq:H2JConFock_plus}
    \\
    \nonumber
    &+&
    g_2\sqrt{n_2+1}\vert n_1,n_2+1;-1 \rangle
   \,, 
\end{eqnarray}
for the two different values of the atomic excitation.
Apart from a term proportional to the initial vector, there are only two new states generated by the action of the Hamiltonian. Since this is true for any Fock state, it implies that all the components of the FSL have the structure of a chain where the sites are associated with Fock basis elements. Since a charge sector labeled by $q_2$ contains $2q_2+1$ states, the components of the FSL are chains with size increasing with the charge value. Given the charge sector $q_2$, the endpoints of the FSL represent the states $\ket{q_2,0;-1}$ and  $\ket{0,q_2;-1}$.

The actions \eqref{eq:H2JConFock_minus} and \eqref{eq:H2JConFock_plus} mean that the Hamiltonian is tridiagonal in the Fock basis restricted to any charge sector. This establishes a connection between the FSL and a Krylov space analysis. Indeed, we can immediately conclude that the FSL in each charge sector is the same as the Krylov chain of the model if we choose  one of the states representing the endpoints of the FSL components as the initial state. This means not only that the number of sites on the chains is the same but also that they represent the same quantum states. In these cases, the dynamics on the FSL and on the Krylov chain are identical.
The Lanczos coefficients that govern this dynamics can be read directly from \eqref{eq:H2JConFock_minus}.
If, for instance, we consider the initial state
 $\vert\psi_0\rangle=\vert q_2,0;-1\rangle $, we obtain
 \begin{equation}
 \label{eq:Lanczos_an_JC_N2}
     a_{2k}=-\frac{\Delta}{2}+\omega_1 q_2+k(\omega_2-\omega_1),
     \quad 0\leq k\leq q_2\,,
     \qquad
     a_{2k+1}=\frac{\Delta}{2}+\omega_1\left(q_2-1\right)+k(\omega_2-\omega_1),
     \quad 0\leq k\leq q_2-1\,,
 \end{equation}
 and
  \begin{equation}
  \label{eq:Lanczos_bn_JC_N2}
     b_{2k+1}=g_1\sqrt{q_2-k}\,,
     \qquad
     b_{2k+2}=g_2\sqrt{k+1}\,,
     \qquad
     0\leq k\leq q_2-1\,.
 \end{equation}
\begin{figure}[ht]
    \centering
    \begin{subfigure}{0.48\textwidth}
        \centering
        \includegraphics[width=1.0\textwidth]{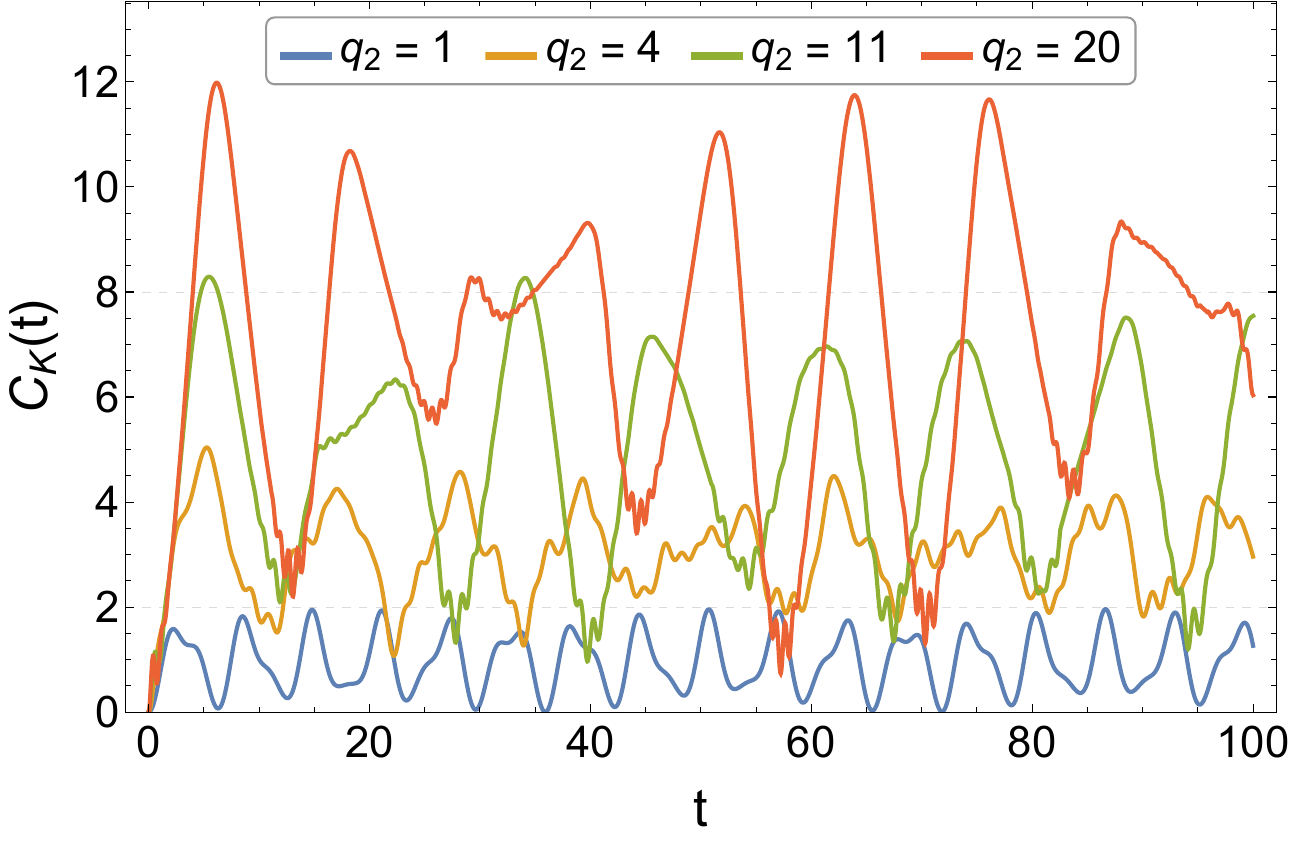}
    \end{subfigure}
    \hfill
    \begin{subfigure}{0.48\textwidth}
        \centering
        \includegraphics[width=1.0\textwidth]{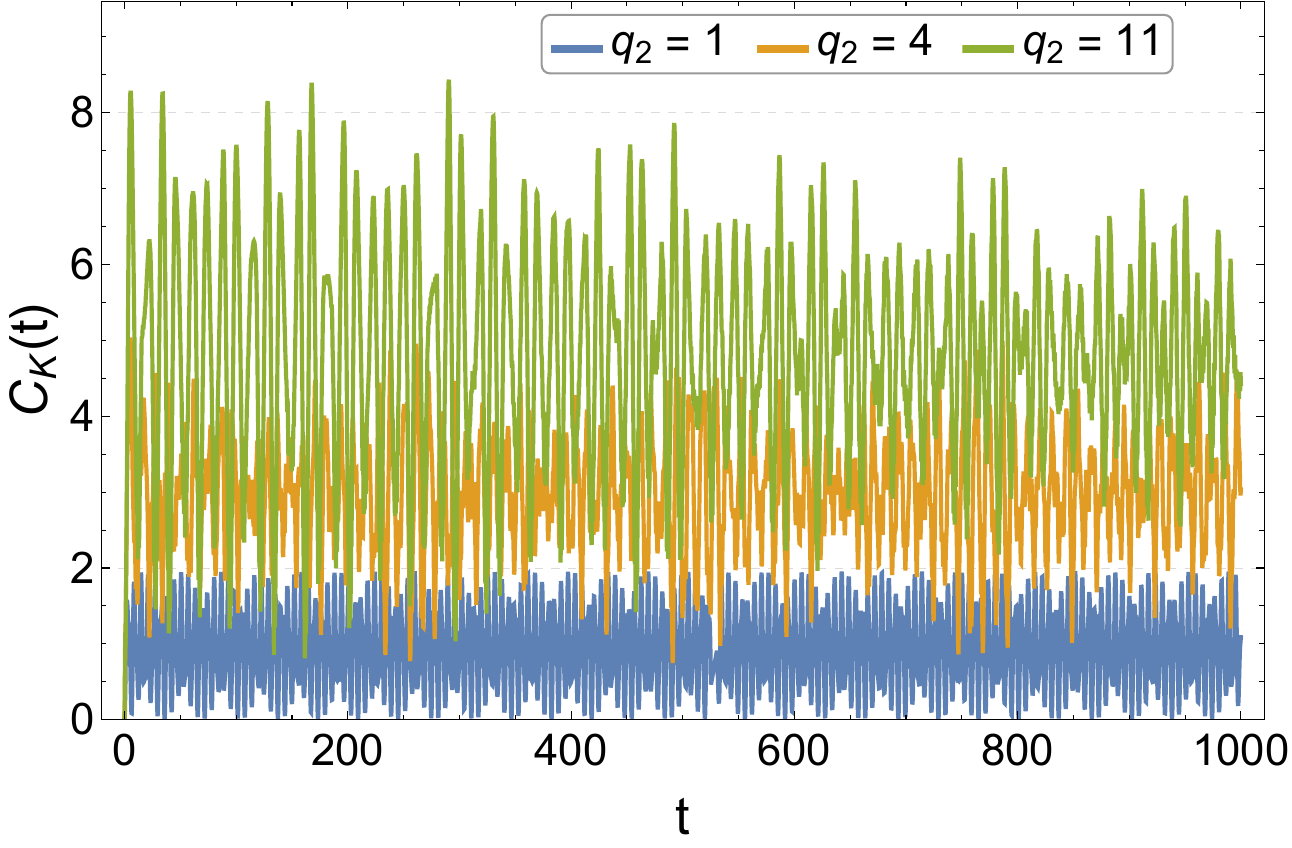}
    \end{subfigure}
    \caption{Krylov complexity of the JC model with Hamiltonian \eqref{eq:JCham_Nmode} with $N=2$ and initial state $\vert\psi_0\rangle=\vert q_2,0;-1\rangle $. The curves are plotted as functions of time and reported for different charge sectors $q_2$. In the right panel, the same curves are plotted for longer times. The parameters in \eqref{eq:JCham_Nmode} are $\Delta=2$, $\omega_1=1$, $\omega_2=1.4$, $g_1=0.9$, and $g_2=1.1$  for all the reported curves.}
\label{fig:JC_N2_complexity}
\end{figure}
The evolution of the hopping equation \eqref{eq:evolutionKamplitude} with Lanczos coefficients given by \eqref{eq:Lanczos_an_JC_N2} and \eqref{eq:Lanczos_bn_JC_N2} can be solved numerically sector by sector. The time evolving amplitude gives the Krylov complexity per \eqref{eq:def_spreadcomplexity}. In Fig.\,\ref{fig:JC_N2_complexity}, we report the evolution of $C_K$ as a function of time for different choices of the charge sector $q_2$. As expected, the values around which $C_K$ oscillates increase with $q_2$, corresponding to an increasing Krylov space dimensionality.
In the right panel, the same curves are plotted over longer times. The oscillations persist at late times and remain incoherent.
At early times, the curves obtained for different values of $q_2$ collapse onto one
another, the agreement improving as $q_2$ increases. This can be understood from the
staggering of the Lanczos coefficients $b_k$ in \eqref{eq:Lanczos_bn_JC_N2}. In
dimerized chains, the group velocity is controlled by the weaker of the two hopping
parameters \cite{Asboth:2016gfa}. For $q_2$ large enough, $b_{2k+2}$ is the weaker coupling throughout the
portion of the chain explored at early times, so that the group velocity, and hence
the early-time growth of the Krylov complexity, becomes independent of $q_2$, giving
rise to the collapse.

It is insightful to discuss what happens in the large-detuning limit, where the Hamiltonian becomes \eqref{eq:multimodeJC_largedet} with $N=2$. Due to the presence of two modes only, in each charge sector and for both values of the atomic excitation, this dynamics occurs on a one-dimensional FSL, whose endpoints represent the states $|n_1+n_2,0;\sigma\rangle$ and $|0,n_1+n_2;\sigma\rangle$.  Indeed, each charge sector is labeled by the sum of the photon number in each mode, i.e. $n_1+n_2$, contains $n_1+n_2+1$ states, and is independent of the polarization $\sigma$ (strictly speaking, there are two charge sectors for any value of $n_1+n_2$, labeled by $\sigma=\pm 1$). As for general choices of parameters, also in the large-detuning regime the dynamics on the FSL and the Krylov chain dynamics of an initial state being at the edges of the FSL are identical, i.e. the Krylov basis of this dynamics is exactly the Fock basis.

The on-site energies and the hopping parameters along the chain, i.e. the Lanczos coefficients in the Krylov space language,  now have a different form and read
\begin{equation}
    a_k=\frac{\Delta}{2}\,\sigma
+\frac{(1+\sigma)\left(g_1^2+g_2^2\right)}{2\Delta}
+\left(\omega_1+\frac{\sigma g_1^2}{\Delta}\right)(n_1+n_2)
+k\left(\frac{\sigma\left(g_2^2-g_1^2\right)}{\Delta}+\omega_2-\omega_1\right),
\quad 0\le k\le n_1+n_2\,,
    \end{equation}
    and
    \begin{equation}
     b_k=\frac{ g_1g_2}{\Delta}\sqrt{k(n_1+n_2-k+1)}\,,\qquad 1\leq k\leq n_1+n_2\,.
\end{equation}
The peculiar linear growth of $a_k$ and the form of $b_k$ allow us to recognize a pattern that helps to solve the dynamics analytically. Indeed, these coefficients are the ones that describe the solvable dynamics induced by an evolution Hamiltonian 
\begin{equation}
  H= \alpha(J_++J_-)+\gamma J_0 \,,\quad \alpha=\frac{g_1 g_2}{\Delta}\,,\quad \gamma=\frac{\sigma}{\Delta}(g_2^2-g_1^2)+\omega_2-\omega_1\,, 
\end{equation}
 where $J_0$ and $J_\pm$ are the $\mathfrak{su}(2)$ generators, acting on an initial highest-weight state of the irreducible representation of dimension $2j+1=n_1+n_2+1$.
 This Krylov-space dynamics was solved in \cite{Caputa:2021sib,Balasubramanian:2022tpr} and we refer to these works for all the computational details. 
For our purposes, it is enough to stress that we can obtain an analytical formula for the Krylov complexity, which reads
\begin{equation}
\label{eq:complexitySU2}
C_K(t)=\frac{n_1+n_2}{\mathcal{D}}\sin^2\left(\frac{g_1 g_2 \sqrt{\mathcal{D}}}{\Delta} t\right)\,,\qquad
\mathcal{D}\equiv 1 + \frac{\left[(g^2_1 - g^2_2)\sigma + \Delta(\omega_1 - \omega_2)\right]^2}{4 g_1^2 g_2^2}\,.
\end{equation}
In this formula, the oscillations are due to the finite dimensionality of the Krylov space in any charge sector, while the coherence of such oscillations, as opposed to the ones observed in Fig.\,\ref{fig:JC_N2_complexity}, is due to the effective $\mathfrak{su}(2)$ symmetry emerging in the large-detuning regime.

From a physical point of view, these results are particularly interesting. The Hilbert space of the full system is infinite dimensional, since the Hamiltonian involves two oscillators, each associated with an infinite-dimensional Hilbert space. However, the Hilbert space within each fixed-charge sector is finite dimensional, with dimension $2q_2+1$ (or $n_1+n_2+1$ in the large-detuning regime). This has an important consequence.
If the initial state is a superposition of vectors belonging to {\it finitely many} charge sectors, then the dynamics necessarily takes place within a finite-dimensional Krylov space. On the other hand, if the initial state has support on infinitely many charge sectors nothing prevents, in principle, the dynamics from exploring an infinite-dimensional Krylov space.
This distinction can be rephrased by saying that, in the first case, the Krylov complexity cannot grow without bound as $t\to\infty$, whereas in the second case such an unbounded growth may occur. It would therefore be interesting to derive a bound on the Krylov complexity that could clarify whether an unbounded late-time growth can actually arise in some situations or is instead fundamentally ruled out. Using the language from \cite{Caputa:2025mii,Caputa:2025ozd}, the Krylov complexities in Fig.\,\ref{fig:JC_N2_complexity} and in \eqref{eq:complexitySU2}, which refer to fixed charge sectors, are symmetry-resolved Krylov complexities. We hope that tools from this context could be helpful in the future in addressing these questions.

In all the cases considered above, the Krylov chain is identical to the FSL. This is due to the choice of the initial state of the Krylov dynamics as one of the endpoints of the FSL. If the initial state is a generic state in the charge sector, the Krylov dynamics occurs, as always, on a chain, but the states represented by the vertices of the chain are not the Fock states, but linear combinations thereof. Thus, although the Krylov chain and the FSL continue to have the same dimensionality as lattices, they differ in general. In the next subsection, we will discuss examples where the FSL is no longer a chain, but a higher-dimensional lattice. This constitutes an even more evident distinction between the two structures, which deserves a detailed analysis.

\subsection{Three-mode JC model}\label{subsec:3mode_JC}

One of the most crucial differences between the FSL and the Krylov chain is that while the dynamics in the Krylov basis can always be represented on a one-dimensional discrete geometry whose features, such as its length, depend on the choice of Hamiltonian and initial state, the FSL structure, including its dimensionality and connectivity, depends on the model.
An example where this difference emerges clearly is the three-mode JC model.

\subsubsection{The Fock state lattices}

\begin{figure}[t!]
    \centering
    \begin{subfigure}{0.46\textwidth}
        \centering
        \includegraphics[width=1.0\textwidth]{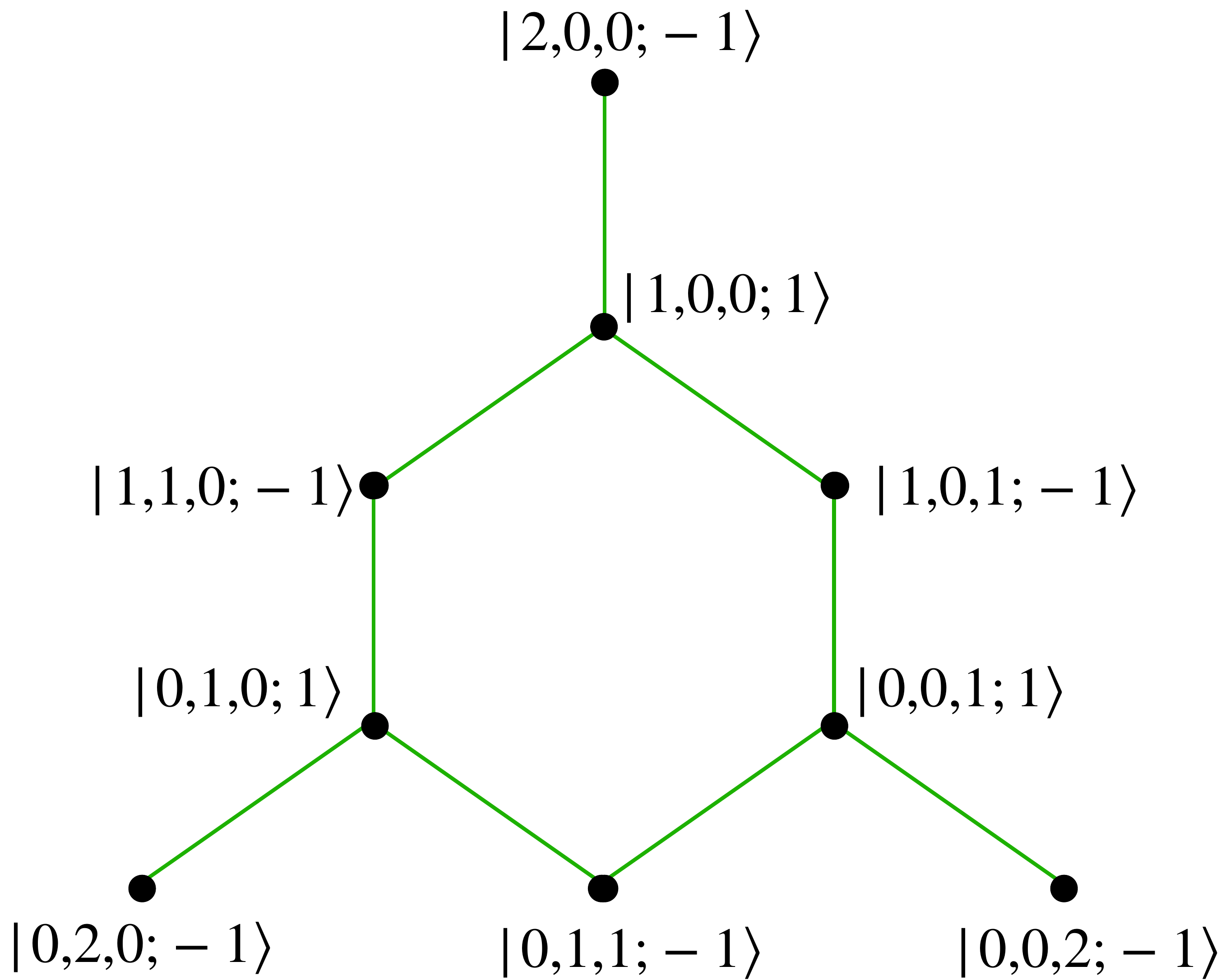}
    \end{subfigure}
    \hfill
    \begin{subfigure}{0.51\textwidth}
        \centering
        \includegraphics[width=1.0\textwidth]{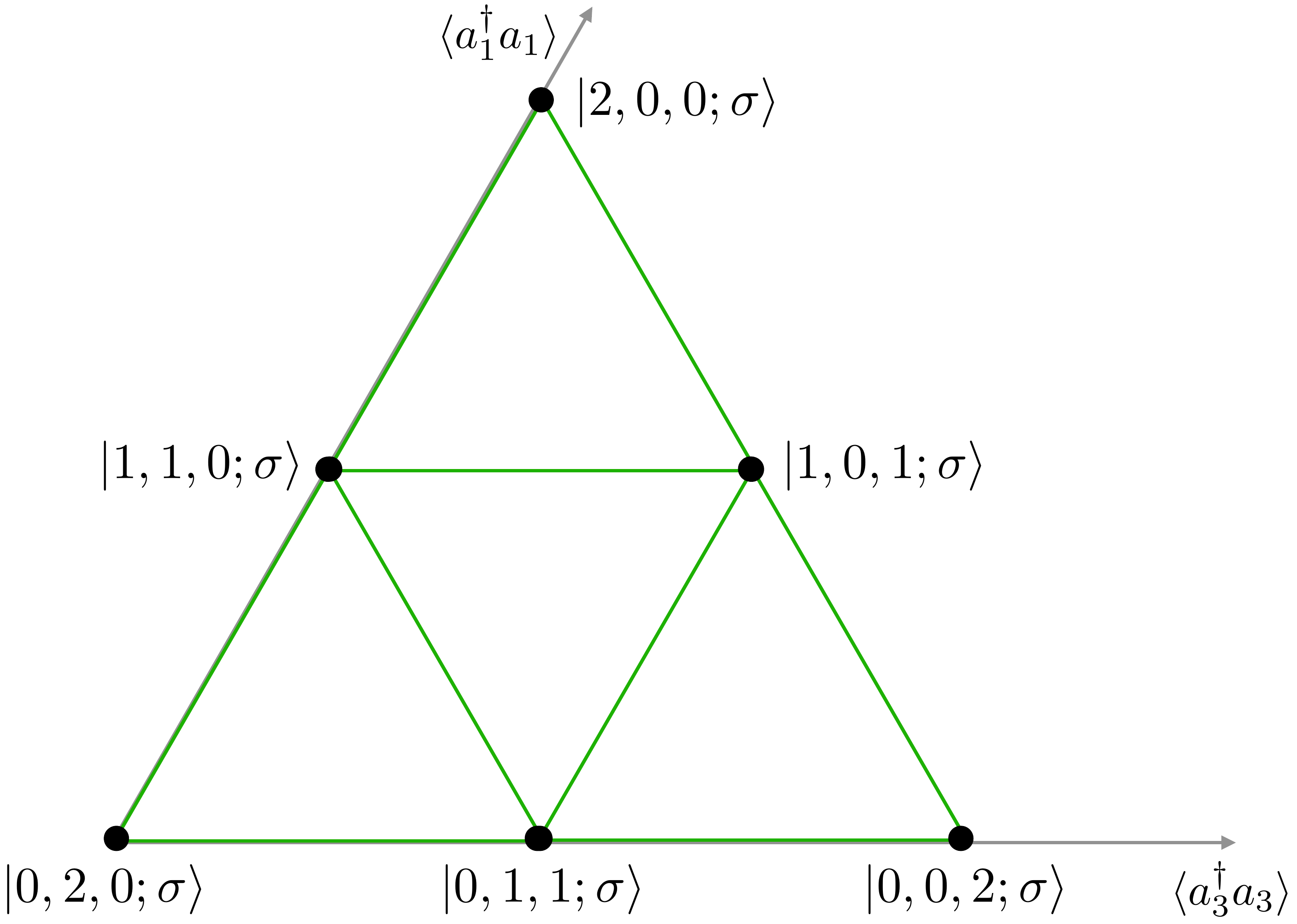}
    \end{subfigure}
    \caption{FSL of the three-mode JC model in a fixed charge sector. In the left panel, the FSL is obtained from the Hamiltonian \eqref{eq:JCham_Nmode} with $N=3$ in the charge sector with $q_3=2$, while the FSL in the right panel is constructed from \eqref{eq:multimodeJC_largedet} in the large-detuning limit and in the sector labeled by $\mathcal{N}_3=2$.}
\label{fig:JC_N3_FSLs}
\end{figure}

By studying the Hamiltonian  \eqref{eq:JCham_Nmode} with $N=3$, we observe that the coordination number of the resulting FSL is not strictly two, but varies between two and three depending on the site (excluding the corners) \cite{Saugmann:2022qzh}.
Considering all the charge sectors leads to an infinite set of two-dimensional FSLs labeled by the charge $q_3$, that is the eigenvalue of  \eqref{eq:JCmodeQ_Nmode}; an example of this lattice, for the charge sector labeled by $q_3=2$, is reported in the left panel of Fig.\,\ref{fig:JC_N3_FSLs}.
All the FSL components have finitely many sites, a rotational symmetry $\mathbb{Z}_3$ around the central plaquette, and the three most external sites are always associated with the states $|q_3,0,0;-1\rangle$, $|0,q_3,0;-1\rangle$ and $|0,0,q_3;-1\rangle$.

An analogous situation occurs in the large-detuning limit. The Hamiltonian \eqref{eq:multimodeJC_largedet} with $N=3$ allows us to construct an infinite series of disconnected FSL components. These are finite-size triangular lattices labeled by the possible total numbers of photons $\mathcal{N}_3=n_1+n_2+n_3$. Since the spin variable is decoupled in this limit, for any charge sector, there are two copies of the same FSL associated with the two values of the atomic excitation labeled by $\sigma$. An example of FSL in the large-detuning regime is shown in the right panel of Fig.\,\ref{fig:JC_N3_FSLs} for $\mathcal{N}_3=2$.
Due to the triangular structure, all the FSL components have a discrete rotational $\mathbb{Z}_3$ symmetry and the external vertices of the lattices represent $|\mathcal{N}_3,0,0;\sigma\rangle$, $|0,\mathcal{N}_3,0;\sigma\rangle$ and $|0,0,\mathcal{N}_3;\sigma\rangle$, with $\sigma=\pm 1$.

A natural question that arises when we study the evolution of a model from the FSL perspective concerns how we track the dynamics represented on the lattice. 
Within the Krylov space approach, a unique answer can be given for every model, i.e. we follow the quantum dynamics of the particle hopping on the Krylov chain and track its average position, i.e. the Krylov complexity, or higher moments. 
On the other hand, the FSL structure varies on a case-by-case basis and therefore, finding a unifying paradigm and a corresponding set of quantities to probe the dynamics of the system on this emergent lattice is less natural. In what follows, we present some approaches for this task, applying them to the evolution of the three-mode JC model. In addition, we compare them to each other and with the corresponding Krylov space dynamics.

\subsubsection{Quantum graph distance}
\label{subsec:graphdistance}

We first consider the notion of a graph distance in the FSL. For illustration purposes, it is more convenient to start from the three-mode JC model in the large-detuning limit, with triangular FSL, as the one represented in the right panel of Fig.\,\ref{fig:JC_N3_FSLs}. The triangular lattices identify two natural axes, parameterized by the average occupation numbers of two out of the three bosonic modes. Conventionally and in agreement with Fig.\,\ref{fig:JC_N3_FSLs}, we choose the axes to be  parametrized by the expectation values of $a_3^\dagger a_3$ and $a_1^\dagger a_1$, so that the origin of the axes is represented by $\ket{0,\mathcal{N}_3,0;\sigma}$. By the $\mathbb{Z}_3$ symmetry of the lattice, the axes and the corresponding origin can be moved by cyclically permuting the three modes.

Consider the initial state coinciding with the state representing the origin of the axes, i.e., $\ket{\psi_0}=\ket{0,\mathcal{N}_3,0;\sigma}$, and let it evolve via \eqref{eq:multimodeJC_largedet} with $N=3$. To capture at the quantum level the number of steps on the FSL graph taken by $|\psi(t)\rangle$ along the two chosen axes, one can naturally define the graph distance of the evolving state from the initial state as
\begin{equation}
\label{eq:graphdistance}
    d_\mathrm{G}(t)\equiv
\bra{\psi(t)}a_1^\dagger a_1\ket{\psi(t)}
    +
\bra{\psi(t)}a_3^\dagger a_3\ket{\psi(t)}=
\mathcal{N}_3- \bra{\psi(t)}a_2^\dagger a_2\ket{\psi(t)}\,.
\end{equation}
 If we take the initial state to be one of the corners of the FSL, $d_\mathrm{G}(t)$ can be adapted by replacing $a_2^\dagger a_2$ in the last expression with the number operator of the only occupied mode in the initial state. By construction, whenever $\ket{\psi(t)}$ is equal to a Fock state,  $d_\mathrm{G}(t)$ takes exactly the integer value given by the minimum number of steps on the FSL graph that separate the Fock state from the initial state. More generally, $d_\mathrm{G}(t)$ encodes at quantum level the average number of graph steps taken during the dynamics.
Given that the expectation values of number operators are restricted to the interval $[0,\mathcal{N}_3]$, we have the constraint $0\le  d_\mathrm{G}(t)\le \mathcal{N}_3$ on the graph distance.

In the sector with $\mathcal{N}_3=1$, the FSL is made up of only three vertices, and there is an interesting connection between $d_\mathrm{G}(t)$ and the Krylov complexity of the evolution. Indeed, in this sector, $d_\mathrm{G}(t)=1-|R(t)|^2$, where $R(t)\equiv\bra{\psi(t)}\psi_0\rangle$ is the return amplitude, which is known to contain all the information on the Krylov space dynamics \cite{Parker:2018yvk, Balasubramanian:2022tpr} and therefore on the Krylov complexity.

The notion of \eqref{eq:graphdistance} as a graph distance is strictly tied to the choice of the initial state as one of the corners of the FSL. By taking any other initial state, \eqref{eq:graphdistance} does not have the same interpretation anymore and a generalization to this case requires more work. We comment on possible extensions of \eqref{eq:graphdistance} which allow for arbitrary initial states in Appendix \ref{app:Wasserstein}.

The definition \eqref{eq:graphdistance} also depends on the precise structure of the FSL, for instance, via the definition of the axes. Indeed, one of the inconvenient aspects of working with this class of FSL distances is the strong dependence on the FSL properties. To illustrate this point, we aim to define a graph distance for the evolution of the three-mode JC model away from the large-detuning regime, i.e. when the dynamics occurs on the FSL in the left panel of Fig.\,\ref{fig:JC_N3_FSLs}. In particular, we assume that the initial state is one of the states at the corners of the FSL component associated with the charge sector $q_3$, i.e. $|q_3,0,0;-1\rangle$, $|0,q_3,0;-1\rangle$ or $|0,0,q_3;-1\rangle$, and we let it evolve with the Hamiltonian in \eqref{eq:JCham_Nmode} with $N=3$. If the initial state has a non-vanishing average of the occupation number $a_j^\dagger a_j$, which decreases as one moves away from the initial state, the number of steps taken on the graph can be quantified by
\begin{equation}
\label{eq:graphdistance_2ndexample}
d_\mathrm{G}(t)= 2 q_3-2\bra{\psi(t)}a_j^\dagger a_j\ket{\psi(t)}-\frac{\bra{\psi(t)}\sigma^{(z)}\ket{\psi(t)}+1}{2}  \,,
\end{equation}
where the constant term fixes the offset so that $d_\mathrm{G}(0)=0$, and the last one keeps track of the change in the atomic excitation along the graph.
Also in this example, it is straightforward to verify that whenever $\ket{\psi(t)}$ is a Fock state, $d_\mathrm{G}(t)$ returns the integer value equal to the minimum number of steps on the graph needed to reach that Fock state from the initial one. This makes \eqref{eq:graphdistance_2ndexample} a well-defined quantum generalization of the graph distance between the initial and evolving state.
This graph distance takes values in the range $0\leq d_\mathrm{G}(t) \leq 2 q_3$, which differs from the range of \eqref{eq:graphdistance}, and that difference precisely reflects the distinct nature of the two dynamics. In the large-detuning regime the spin does not explicitly mediate the interaction between photons: the connectivity of the FSL is thereby enhanced and the larger number of available paths reduces $d_\mathrm{G}(t)$. Away from the large-detuning limit, on the contrary, the participation of the spin in the dynamics imposes additional constraints, so that fewer paths are available and $d_\mathrm{G}(t)$ is correspondingly larger.
This explicit example shows how strongly the definition of $d_\mathrm{G}(t)$ depends on the structure of the FSL. The analysis of other models, including the extension to the generic $N$-mode JC model, therefore has to be carried out on a case-by-case basis.

\subsubsection{Euclidean distance on the FSL}
\label{subsec:Euclidean}

To define an alternative distance on the FSL between the evolving and the initial state, we preliminarily observe that the state $\ket{\psi(t)}$ identifies a time trajectory in the space enclosed in the FSL. For example, the trajectory associated with $\ket{\psi(t)}$ being a state evolving via the Hamiltonian \eqref{eq:multimodeJC_largedet} and starting from a state in the charge sector  labeled by $\mathcal{N}_3$ will lie in the triangle identified by $|\mathcal{N}_3,0,0;\sigma\rangle$, $|0,\mathcal{N}_3,0;\sigma\rangle$ and $|0,0,\mathcal{N}_3;\sigma\rangle$.
Given the allowed region for the trajectory, which has the same dimensionality $D$ as the space where the FSL is embedded, it is natural to identify a set of orthogonal axes defining Euclidean coordinates $\mathbf{x}=(x_1,\dots,x_D)^{\rm t}\in\mathbb{R}^D$. The trajectory can be, therefore, represented by a vector $\mathbf{x}(t)$ in these coordinates and a distance between the initial and the evolving state is given by the Euclidean distance between $\mathbf{x}(t)$ and $\mathbf{x}(0)$, i.e.
\begin{equation}
    d_\mathrm{E}(t)\equiv
    \sqrt{\sum_{j=1}^D[x_j(t)-x_j(0)]^2}
\,.
\label{eq:Euclideandist}
\end{equation}
It is convenient to illustrate also this example via the FSL in the right panel of Fig.\,\ref{fig:JC_N3_FSLs}, namely for the large-detuning three-mode JC dynamics of a state in a given charge sector $\mathcal{N}_3$.
This FSL is a planar graph, so that $D=2$.
It is natural to identify the two Euclidean coordinates as
\begin{equation}
x_1(t)=\frac{1}{2}\bra{\psi(t)}a_3^\dagger a_3-a_2^\dagger a_2\ket{\psi(t)}\,,\qquad x_2(t)=\frac{\sqrt{3}}{2}\bra{\psi(t)}a_1^\dagger a_1\ket{\psi(t)}\,.
     \label{eq:coordinateN3n1}
 \end{equation}
 The prefactors of the coordinates, in principle arbitrary, are chosen in such a way that adjacent sites have distance equal to one. In this way, the corresponding $d_\mathrm{E}$ is comparable with $d_\mathrm{G}$, which has the same property.
 Once \eqref{eq:coordinateN3n1} is plugged into \eqref{eq:Euclideandist}, these coordinates give the distance $ d_\mathrm{E}(t)$ for this specific model.
Differently from \eqref{eq:graphdistance}, the general distance \eqref{eq:Euclideandist} is defined for any initial state belonging to a given FSL component, i.e. to a charge sector, and not necessarily being a Fock state. Moreover it is upper-bounded by the maximal possible Euclidean distance  in the region allowed to the trajectory, i.e. $ d_\mathrm{E}(t)\leq \mathcal{N}_3$.

The two Euclidean coordinates to plug into \eqref{eq:Euclideandist} can be defined also for the FSL in the left panel of Fig.\,\ref{fig:JC_N3_FSLs} and the dynamics occurring on it. We have
\begin{equation}
\label{eq:coordinateN3hex}
    x_1(t)=\frac{\sqrt{3}}{2}\,\bra{\psi(t)}a_3^\dagger a_3-a_2^\dagger a_2\ket{\psi(t)}\,,
    \qquad
    x_2(t)=\bra{\psi(t)}a_1^\dagger a_1-\frac{1}{2}\left(a_2^\dagger a_2+a_3^\dagger a_3\right)\ket{\psi(t)}\,,
\end{equation}
where, again, the prefactors guarantee that every pair of sites connected by a single edge has a distance $d_\mathrm{E}$ equal to one. Notice that none of these coordinates depends on the expectation value of the spin variable $\sigma^{(z)}$. This is due to the fact that these coordinates are always evaluated in a sector with fixed charge $q_3$, which introduces a relation among the expectation values of the three modes and the atomic excitation, which therefore is not independent and does not need to enter in the coordinates \eqref{eq:coordinateN3hex}.

As an alternative to measures of distance on the FSL between the initial and evolving state, there exist different ways of quantifying the spread of the evolving state on the FSL or, in mathematical terms, on the Fock basis. An example is given by the participation entropy. Given the amplitude of the evolving state on the Fock basis
\begin{equation}
 \label{eq:FSLamplitude}  \psi_{n_1,n_2,n_3}^{(\sigma)}(t)=\langle n_1,n_2,n_3;\sigma|\psi(t)\rangle \,,
\end{equation}
the FSL participation entropy is given by
\begin{equation}
\label{eq:FSLentropy}
   S_\mathrm{\tiny FSL}(t)=-
   \sum_{\mathrm{\tiny FSL}}p_{n_1,n_2,n_3}^{(\sigma)}(t)\ln p_{n_1,n_2,n_3}^{(\sigma)}(t)\,,
   \qquad
   p_{n_1,n_2,n_3}^{(\sigma)}(t)\equiv\vert\psi_{n_1,n_2,n_3}^{(\sigma)}(t) \vert^2\,,
\end{equation}
where the sum is done over all the states belonging to the same FSL component as the initial state. This quantity is the FSL-counterpart of the K-entropy \eqref{eq:Kentropy}, similarly defined with respect to the Krylov basis, and, therefore, measuring the spread along the Krylov chain \cite{Balasubramanian:2022tpr}.

In what follows, we study the distance measures and the spread quantifier introduced in the previous subsections and we compare them with the same quantities defined on the Krylov chain. We will focus for simplicity on the three-mode JC model in the large-detuning regime.

\subsubsection{Numerical analysis}

We restrict our analysis to systems in finite-dimensional charge sectors. Thus, all the studied quantities will exhibit oscillations in time. To mitigate the effect of these oscillations, we plot and compare the time-averaged quantities. Given a quantity $O(t)$, its time-average is defined as
\begin{equation}
    \bar{O}(T)=
    \frac{1}{T}\int_0^T O(t)dt
    \,.
\end{equation}
In the limit $T\to\infty$, this definition gives the average value around which $O(t)$ oscillates. For $T$ finite, but large enough, $\bar{O}(T)$ exhibits an initial growth in $T$ followed by a saturation to the average value of interest (with possible residual small oscillations around it).

\begin{figure}[t]
\centering
\begin{minipage}[c]{0.15\textwidth}
  \centering
  \footnotesize
  \setlength{\tabcolsep}{4pt}
  \renewcommand{\arraystretch}{1.3}
  \begin{tabular}{cccc}
  \toprule
   & $(g_1,g_2,g_3)$ & $(\omega_1,\omega_2,\omega_3)$ & $\ket{\psi_0}$ \\
  \midrule
  \clr{red}       & $(1.00,\,1.40,\,1.00)$ & $(1.00,\,0.90,\,1.10)$ & $\ket{3,0,0;\,1}$ \\
\clr{blue}      & $(1.20,\,0.90,\,1.00)$ & $(1.00,\,1.20,\,1.20)$ & $\ket{0,0,3;\,1}$ \\
\clr{darkgreen} & $(0.70,\,0.70,\,0.90)$ & $(1.00,\,0.80,\,1.30)$ & $\ket{0,3,0;\,1}$ \\
\clr{black}     & $(1.30,\,0.90,\,0.90)$ & $(1.00,\,1.00,\,1.00)$ & $\ket{3,0,0;\,1}$ \\
  \bottomrule
  \end{tabular}
\end{minipage}\hfill
\begin{minipage}[c]{0.48\textwidth}
  \centering
  \includegraphics[width=\linewidth]{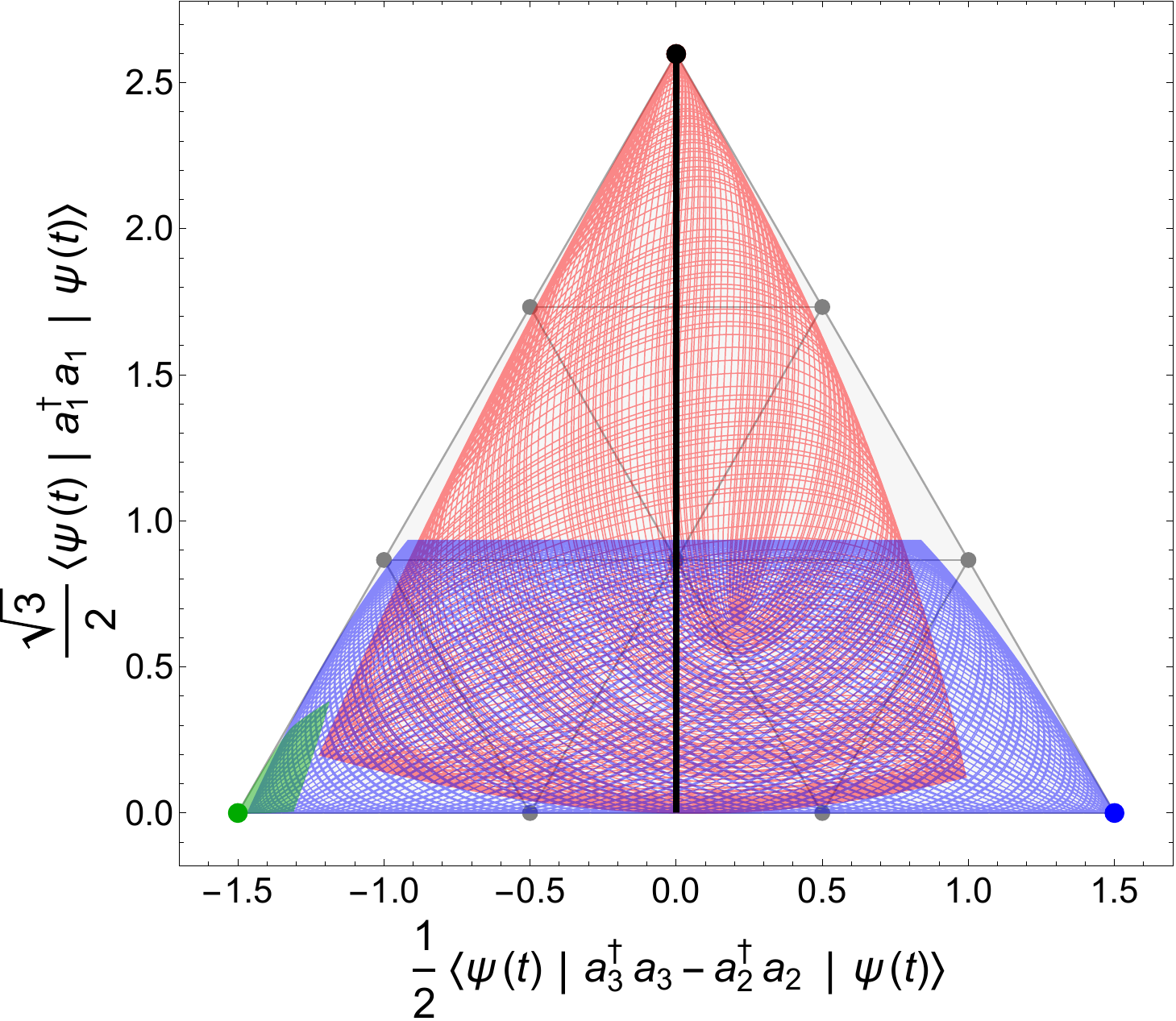}
\end{minipage}
\caption{Left panel: The four dynamics considered in the numerical analysis. The
couplings $g_i$ and the frequencies $\omega_i$ are those entering the
Hamiltonian \eqref{eq:multimodeJC_largedet} with $N=3$; in all cases
$\Delta=10$ and $\sigma=1$. Each dynamics is identified by a color, used
consistently in all the figures below. Right panel: The corresponding
trajectories, drawn in the coordinates \eqref{eq:coordinateN3n1}, inside the
triangular region delimited by the FSL of the $N=3$ JC model in the
large-detuning limit and in the charge sector $\mathcal{N}_3=3$.}
\label{fig:dynamics}
\end{figure}

In the numerical analysis that follows, we study the evolution generated by
\eqref{eq:multimodeJC_largedet} with $N=3$, starting from an initial state
$\ket{\psi_0}$ belonging to the charge sector labeled by $\mathcal{N}_3=3$ . We consider four choices of the parameters of the evolution
Hamiltonian, each with its own initial state. The resulting four dynamics are
summarized in the table in the left panel of Fig.\,\ref{fig:dynamics}, which also fixes the color used to identify
each of them in all the figures below.

We first analyze the trajectories that the four dynamics draw in the region
delimited by the FSL. The trajectories, reported in the right panel of
Fig.\,\ref{fig:dynamics}, are defined by the time-dependent coordinates
\eqref{eq:coordinateN3n1} and correspond to the dynamics listed in the table in
the left panel, according to the color code; the initial states are marked by a
larger dot of the same color. We observe that different dynamics explore
different subregions of the FSL, whose size depends on the parameters of the
evolution Hamiltonian. When these parameters are symmetric under the exchange of
two of the three modes in \eqref{eq:multimodeJC_largedet}, and the initial state
is invariant under the same exchange, the two-dimensional explored region
collapses to a one-dimensional curve. This is the case for the dynamics shown in
black, for which $g_2=g_3$ and $\omega_2=\omega_3$.

\begin{figure}[t]
\centering
\begin{minipage}{0.48\textwidth}
  \centering
  \includegraphics[width=\linewidth]{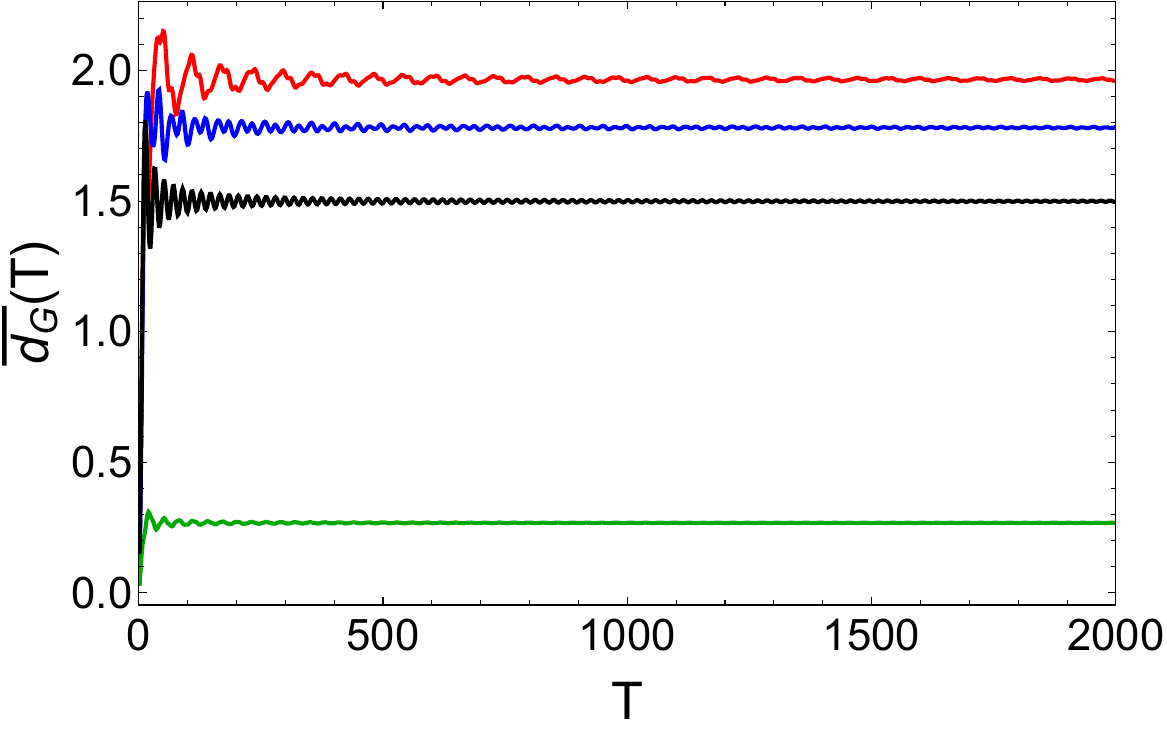}
\end{minipage}\hfill
\begin{minipage}{0.48\textwidth}
  \centering
  \includegraphics[width=\linewidth]{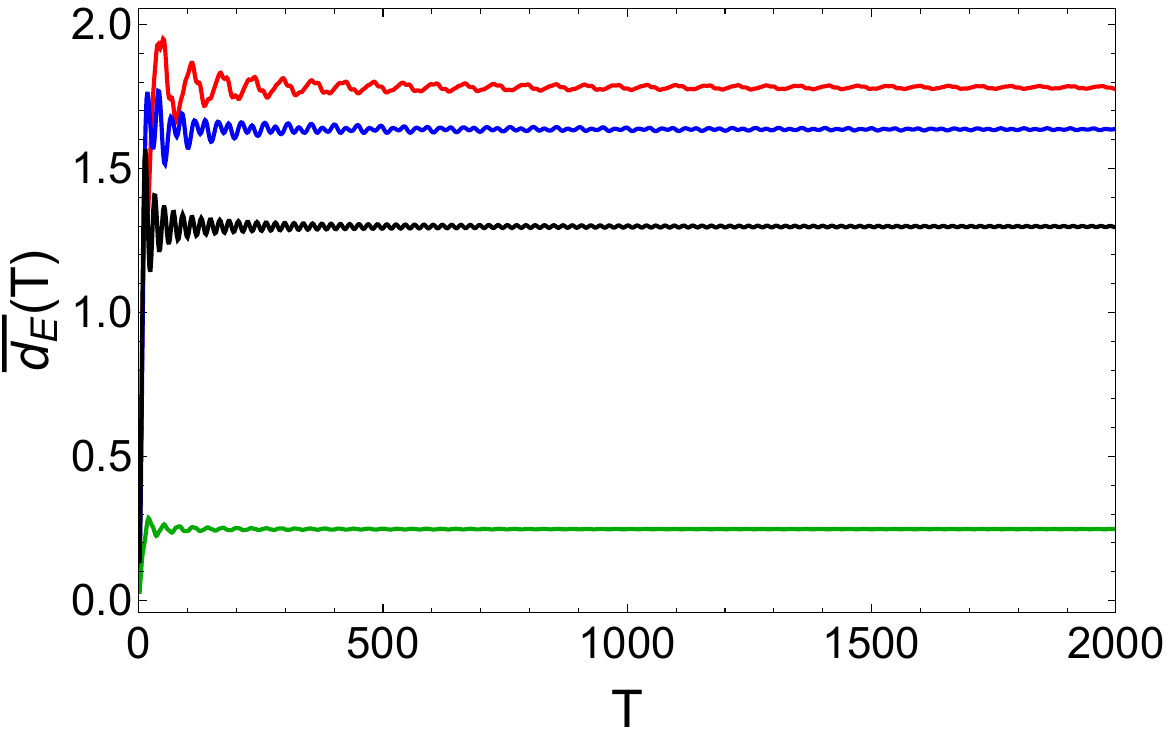}
\end{minipage}

\vspace{\floatsep}

\begin{minipage}{0.48\textwidth}
  \centering
  \includegraphics[width=\linewidth]{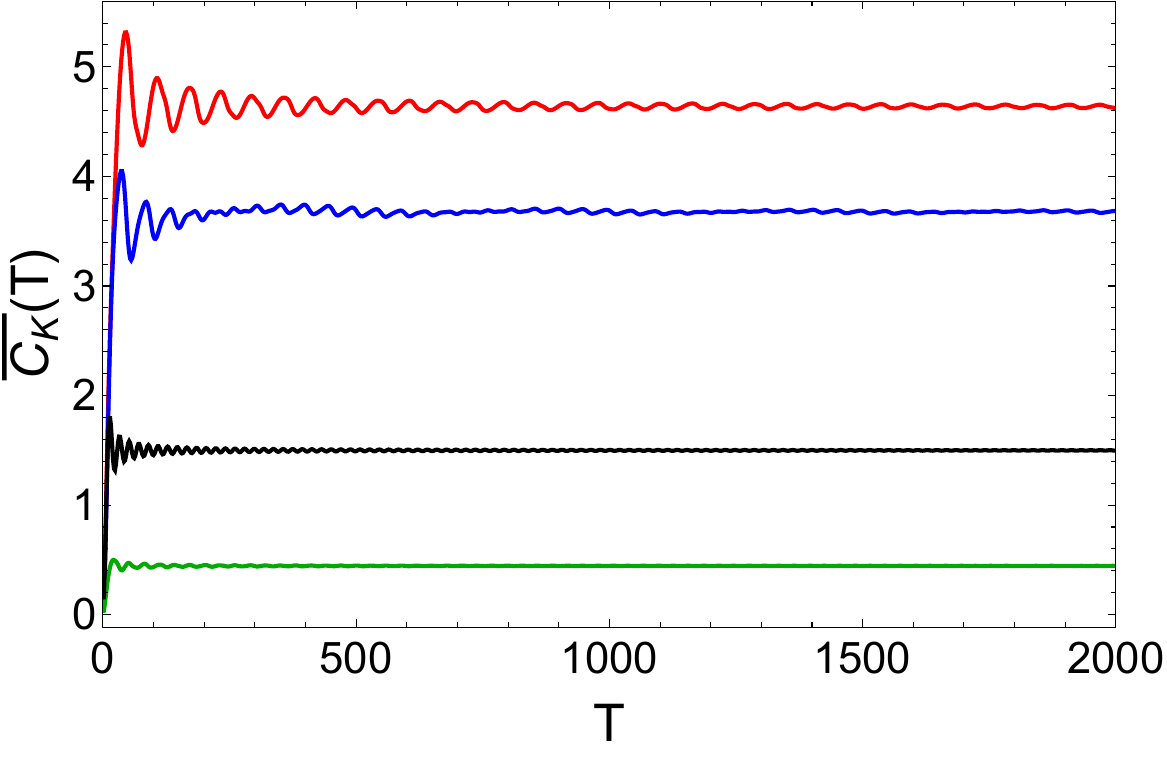}
\end{minipage}
\caption{
Top panels: time-averaged graph distance \eqref{eq:graphdistance} (left)
and Euclidean distance \eqref{eq:Euclideandist}-\eqref{eq:coordinateN3n1} (right) between the evolving
state and the initial one, as functions of $T$, for the evolution generated by
the Hamiltonian \eqref{eq:multimodeJC_largedet} with $N=3$. Bottom
panel: time-averaged Krylov complexity \eqref{eq:def_spreadcomplexity} for the
same time evolution.
In all
three panels, the parameters of the evolution Hamiltonian and the initial states
are those listed in the table of Fig.\,\ref{fig:dynamics}, which also fixes the
color code of the curves.}
\label{fig:timeaverages}
\end{figure}

The regions explored by the four dynamics have different areas. One might expect
larger areas to correspond to larger typical distances between the evolving state
and the initial one. The time-averaged distances we consider, however, are
sensitive not only to which points are visited, but also to how long the dynamics
lingers near each of them. If an evolving state spends most of its time close to
the initial state, its average distance will be small, whichever definition is
used, even if it briefly visits points very far away.

In Fig.\,\ref{fig:timeaverages} we report the time averages of the graph distance
\eqref{eq:graphdistance}, of the Euclidean distance \eqref{eq:Euclideandist}
evaluated with the coordinates \eqref{eq:coordinateN3n1}, and of the spread
complexity \eqref{eq:def_spreadcomplexity}, as functions of $T$.
Although the three panels display different measures, the curves behave in a
similar way; in particular, the late-time saturation values order the four
dynamics in the same way. Comparing with the right panel of
Fig.\,\ref{fig:dynamics}, larger saturation values correspond to larger linear
sizes of the region explored by the trajectories, a feature that all three
quantities capture.
It is worth pointing out that no direct relation holds between these
saturation values and the volumes (areas, in the present two-dimensional case)
of the regions explored on the FSL. The dynamics shown in green explores a region
of non-zero area, whereas the black one is confined to a one-dimensional
manifold; nevertheless, the black curves saturate at considerably larger values
than the green ones in all three panels. This supports the conclusion that what
controls the time-averaged distances is the linear size of the explored region
rather than its volume.

\begin{figure}[t]
\centering
\begin{minipage}{0.48\textwidth}
  \centering
  \includegraphics[width=\linewidth]{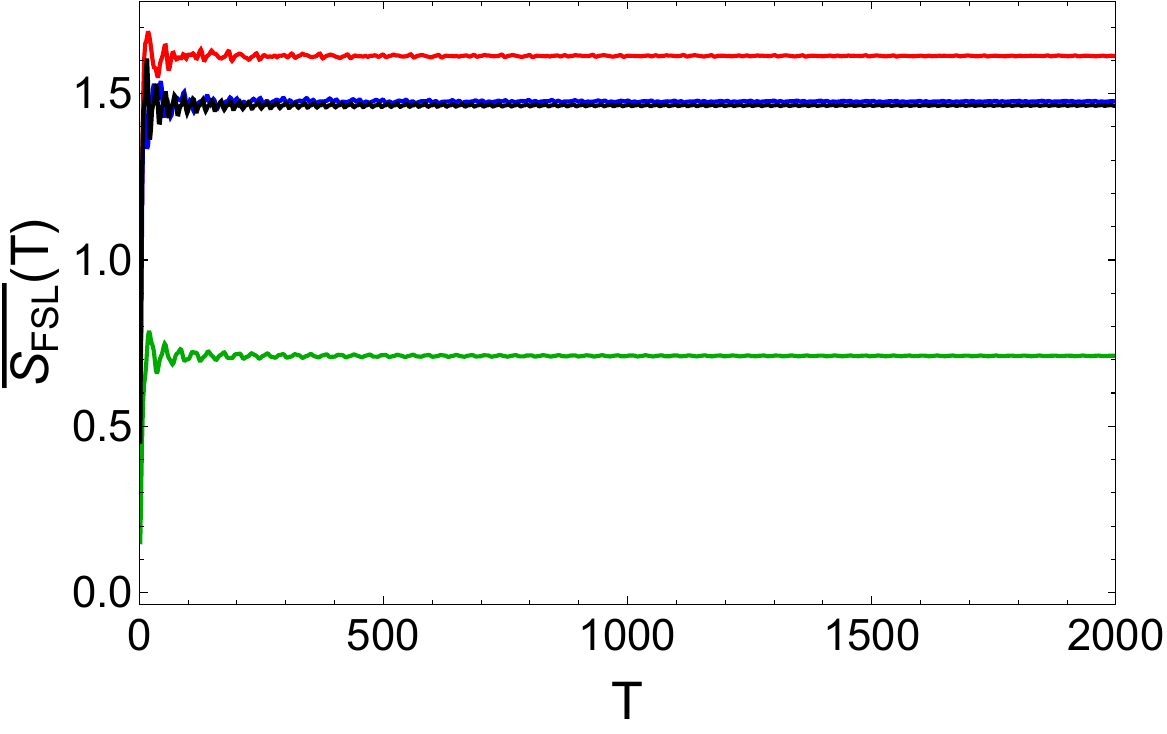}
\end{minipage}\hfill
\begin{minipage}{0.48\textwidth}
  \centering
  \includegraphics[width=\linewidth]{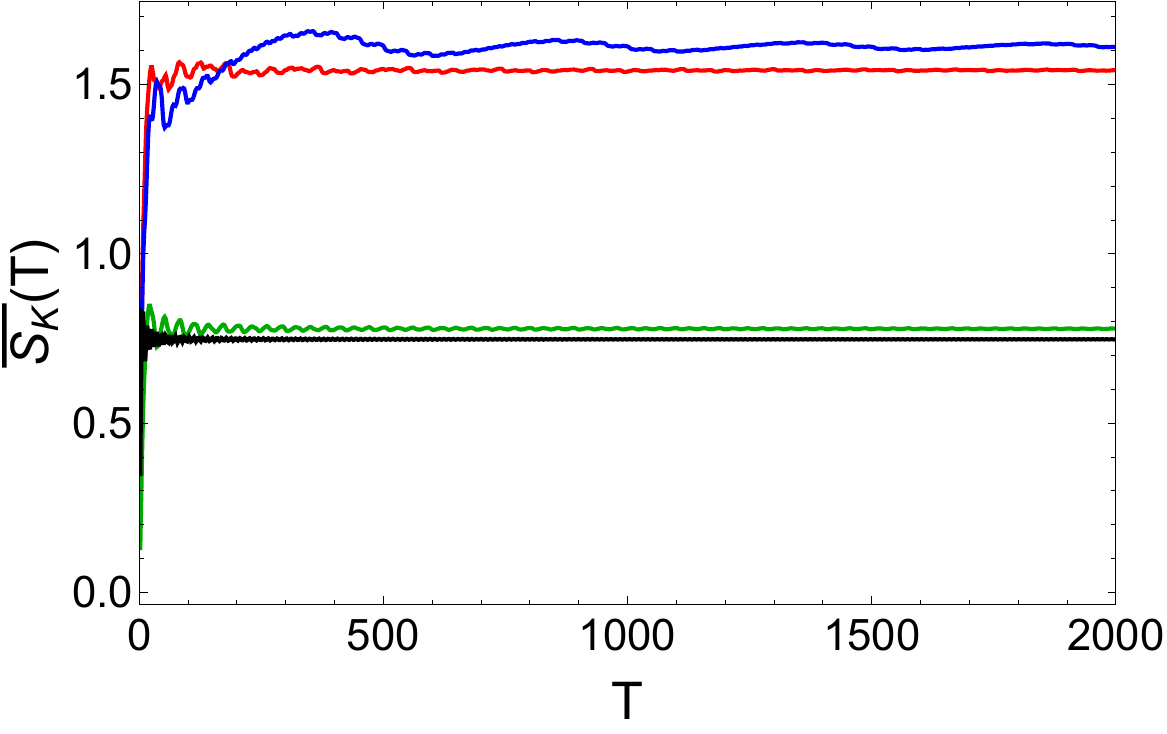}
\end{minipage}
\caption{Time-averaged participation entropy \eqref{eq:FSLentropy} (left) and K-entropy
\eqref{eq:Kentropy} (right), as functions of $T$, for a state evolving under the
Hamiltonian \eqref{eq:multimodeJC_largedet} with $N=3$. In both panels, the
parameters of the evolution Hamiltonian and the initial states are those listed
in the table of Fig.\,\ref{fig:dynamics}, which also fixes the color code of the
curves.}
\label{fig:spread}
\end{figure}

In Fig.\,\ref{fig:timeaverages}, we observe that, for each of the four dynamics,
the graph distance \eqref{eq:graphdistance} is always larger than or equal to
the Euclidean distance \eqref{eq:Euclideandist}. This is in fact a general
property, valid whenever the initial state is one of the corners of the FSL, and
it follows from the identity
\begin{equation}
    d^2_\mathrm{G}(t)-d^2_\mathrm{E}(t)=\bra{\psi(t)}a_i^\dagger a_i\ket{\psi(t)}
\bra{\psi(t)}a_j^\dagger a_j\ket{\psi(t)}\geq0\,,
\end{equation} where $i$ and $j$ label the two
bosonic modes that are not occupied in the initial state. Since both distances
are non-negative, this implies $d_\mathrm{G}(t)\geq d_\mathrm{E}(t)$.

Studying the evolving state from the point of view of other spread measures on the Fock basis, i.e. the participation entropy \eqref{eq:FSLentropy}, and on the Krylov basis, i.e. the K-entropy \eqref{eq:Kentropy}, allows us to identify a few tricky points in the representation of quantum dynamics through the trajectories in Fig.\,\ref{fig:dynamics}.
The outcome of our analysis is shown in Fig.\,\ref{fig:spread}, where we report the time-averaged $S_\mathrm{\tiny FSL}$  (left panel) and K-entropy $S_K$ (right panel) of a state evolving according to the four dynamics in the table in Fig.\,\ref{fig:dynamics}.

The first interesting observation is that different spread measures react
differently to the presence of symmetries in the system. This can be seen by
considering the dynamics represented in black, which enjoys an extra symmetry,
the exchange of the second and third bosonic modes, that the other dynamics do
not have. This symmetry is detected by the Krylov analysis: the dimension of the
Krylov space drops from ten to four, so that the maximal value attainable by
$\bar S_K$ decreases from $\ln 10$ to $\ln 4$, and the black curve indeed
exhibits the smallest saturation value. The FSL analysis, instead, does not
reveal the symmetry: the saturation value of the black curve in the left panel is
almost indistinguishable from that of the blue dynamics and only slightly below
the largest one. The reason is that the uniform distribution over the Fock basis
is itself symmetric under the exchange of the two modes, so that the maximum of
$S_\mathrm{FSL}$ is unaffected. In this sense, $S_\mathrm{FSL}$ detects the
symmetries of the system less sharply than $S_K$.

In general, we observe that the late-time saturation values of the two quantities order
the dynamics differently, that is, for a given dynamics the spread in the
two bases is not the same. A reason behind this fact is that the vectors of one basis can be concentrated with respect to those of the other. This is what explains the
difference between the saturation values of the red and blue curves, for which
the Krylov space has the full dimension of the sector and the previous argument
does not apply.

Finally, we observe that the black curve for $S_\mathrm{FSL}$ has a large
saturation value, which is unexpected given that the corresponding trajectory in
Fig.\,\ref{fig:dynamics} is confined to a one-dimensional manifold. Computing
numerically the probabilities $p_{n_1,n_2,n_3}^{(\sigma)}(t)$ entering
\eqref{eq:FSLentropy}, we find that they are non-negligible also on Fock states
that do not lie along the black segment; indeed, most of the weight sits away
from it. The apparent mismatch between the trajectory and this broad
distribution originates from the exchange symmetry discussed above, which forces
the populations of Fock states related by the exchange of the second and third
modes to be equal at all times. Their contributions to the coordinates
\eqref{eq:coordinateN3n1} therefore cancel exactly, and the average occupation
numbers remain pinned to the black segment even though the state is spread over
the whole sector. The trajectory records only the first moment of the
distribution and is blind to its width, which is instead detected by spread measures in
a given basis. This remark suggests using the representation of the
dynamics in terms of the trajectories \eqref{eq:coordinateN3n1} with some care:
they capture average behaviors, but conceal the spreading of the state over the
basis.

\section{Dicke model} \label{Dicke}

As mentioned briefly in Sec.\,\ref{intro to qom}, the Dicke model describes the interaction of a single mode of the electromagnetic field with $M$ two-level atoms. Unlike the previous models which are integrable, the Dicke model is generally not. Additionally, it exhibits a phase transition in the thermodynamic limit, which was initially associated with the onset of chaos \cite{emary2003chaos}. Recent works provide a different, more fine-grained perspective on the chaotic features of the model, which despite being more pronounced above the phase transition, are mostly viewed as a function of energy rather than an inherent characteristic of the phases \cite{Chavez-Carlos:2018ijc}. 

We will discuss these points through the lens of both the Krylov and the FSL frameworks, putting an emphasis on the transition from the normal to the \textit{superradiant} phase, so named because it describes the collective behavior of the atoms which forces them to produce more radiation than they would had they radiated independently \cite{Dicke_or}. Our interest lies precisely in the thermodynamic limit $M\to\infty$ and across the phase transition, where as we show below, quantities on the FSL and the Krylov space can serve as useful diagnostics of the dynamics. Notably, the Krylov portion of this topic was the subject of the recent work \cite{Zhang:2026ocm}, which however provides mostly numerical evidence to support the sensitivity of Krylov-related quantities to the phase transition. Here we will provide robust analytical expressions which verify this claim. 

As a first step, it is useful to perform a series of transformations that bring the Dicke Hamiltonian \eqref{Dicke Hamiltonian} into a form that exposes the additional symmetry structure that arises in the thermodynamic limit, meaning for a large number of atoms $M\to\infty$. Using the so-called Holstein-Primakoff transformation, which maps the total spin operator $S^\alpha\equiv\sum_{j=1}^M\sigma^{(\alpha)}_j/2$, with $\alpha=x,y,z$, to a bosonic mode
\be
S^z\rightarrow -\frac{M}{2}+ b^\dagger b, \quad S^+=b^\dagger \sqrt{M-b^\dagger b}~,
\ee
the Hamiltonian can be recast in the form (in the thermodynamic limit, to leading order in $M$)
\be \label{HP Hamiltonian}
H_{\rm HP}= \omega a^\dagger a +\Delta b^\dagger b + g(a^\dagger +a)(b^\dagger +b)~.
\ee
We can further define $c_\pm=\frac{a\pm b}{\sqrt{2}}$, which changes the Hamiltonian to
\be
H_{\rm HP}=\frac{\omega+\Delta}{2}\left(c^\dagger_+c_++c^\dagger_-c_-\right)
+\frac{\omega-\Delta}{2}\left(c^\dagger_+c_-+c^\dagger_-c_+\right)
+g\left(c^\dagger_+c_+-c^\dagger_-c_-\right)
+\frac{g}{2}\left[(c^\dagger_+)^2+c_+^2-(c^\dagger_-)^2-c_-^2\right]~.
\label{eq:HPHamiltonian_polar}
\ee
This is the so-called polariton basis, which might look unnecessarily more complicated, but it can be a useful representation especially in the case of resonance $\omega=\Delta$ where the second term vanishes. This can serve as a simplified case in which we can build some intuition that will help us tackle the problem in its full generality.

Before proceeding with our analysis, a caveat is in order. In the superradiant
phase, i.e.\ for $g>\frac{\sqrt{\omega \Delta}}{2}$, the Hamiltonian \eqref{HP Hamiltonian} is unbounded
from below, due to the mode $c_-$ in \eqref{eq:HPHamiltonian_polar} becoming an inverted oscillator (see also \eqref{eq:Ham_Dicke_quadratures}). This feature
originates from the linearization of the Holstein-Primakoff map,
$\sqrt{M-b^\dagger b}\to\sqrt M$, which we performed to obtain
\eqref{HP Hamiltonian} and which is justified as long as
$\langle b^\dagger b\rangle\ll M$. This bound constrains $b$ only, since this mode
descends from the collective spin and therefore inherits its finite-dimensional
Hilbert space. Since, as we will see later,
$\langle b^\dagger b\rangle$ grows in time in the superradiant phase, our
results hold within a time window whose extent increases with the number of
atoms $M$.

\subsection{Krylov dynamics in the resonant regime}
Notably, in the case of resonance $\omega=\Delta$, the Hamiltonian \eqref{eq:HPHamiltonian_polar} decomposes into two $\mathfrak{su}(1,1)$ sectors
\be \label{eq: Hpolariton decomp}
H_{\rm HP} = \underbrace{(\omega+g)\,\cd_+c_+ + \frac{g}{2}({\cd_+}^2 + c_+^2)}_{H_+} + \underbrace{(\omega-g)\,\cd_-c_- - \frac{g}{2}({\cd_-}^2 + c_-^2)}_{H_-}.
\ee
 Clearly, the resonant regime is special as it reduces the dynamics to that of two decoupled oscillators. While the standard references on the topic make implicit use of this structure, this is, to our knowledge, the first time that this symmetry enhancement is described explicitly, and we hope it can lead to further advancements, at least on the theory side.
Each $H_\pm$ has the dynamical symmetry $\su$, generated by
\begin{equation}
K_+^{(j)} = \frac{{\cd_j}^2}{2}, \qquad K_-^{(j)} = \frac{c_j^2}{2}, \qquad K_0^{(j)} = \frac{2\cd_j c_j + 1}{4},
\end{equation}
satisfying $[K^{(j)}_0, K^{(j)}_\pm] = \pm K^{(j)}_\pm$ and $[K^{(j)}_
+, K^{(j)}_-] = - 2K^{(j)}_0$.

Analytic expressions for the Lanczos coefficients and the Krylov complexity are
currently out of reach for arbitrary initial states. To gain some analytical
insight, however, we can study the evolution induced by $H_\pm$ separately,
starting from the state $|0_\pm\rangle$ annihilated by $c_\pm$,
$c_\pm|0_\pm\rangle=0$.
Following the prescription of \cite{Caputa:2021sib,Balasubramanian:2022tpr}, it is straightforward to derive the Krylov complexity associated with each individual symmetry subsector to be
\be
\label{eq:SR_su(11)_comp}
C^{(\pm)}_K(t)= \frac{g^2}{2 \alpha^2_\pm} \sin^2{(\alpha_\pm t)}~,
\ee
where 
\be\label{eq: Dicke phase}
\alpha_\pm=\sqrt{\omega(\omega \pm 2g)}~.
\ee 
This creates an interesting window into the dynamics around the phase transition, which occurs at $\omega=2g$. While the complexity of the $\alpha_+$ mode remains oscillatory for any value of the coupling, its $\alpha_-$ counterpart transitions from bounded oscillations, to quadratic growth at the critical point $g_c=\omega/2$, to exponential growth above it.
More concretely, we can distinguish the three following regimes for the Krylov complexity of the $\alpha_-$ mode
\be
\label{eq:SR_su(11)_comp_regimes}
g<g_c:\,C^{(-)}_K(t)= \frac{g^2}{2\alpha^2_-} \sin^2{(\alpha_- t)}, \quad g\sim g_c:\, C^{(-)}_K(t)\simeq \frac{g^2 t^2}{2}, \quad g>g_c:\, C^{(-)}_K(t)= \frac{g^2}{2\abs{\alpha_-}^2} \sinh^2{(\abs{\alpha_-} t)}~.
\ee

While this result hinges on some well-established techniques relying on symmetry, it constitutes a novel observation regarding the role of Krylov complexity as a witness of the superradiant phase transition. This is an important addition to the discourse surrounding the potency of the Krylov framework in detecting and characterizing quantum phase transitions.
We stress that, as anticipated, the late-time growth of
\eqref{eq:SR_su(11)_comp_regimes} at the critical point and in the superradiant
phase is reliable only as long as the approximations leading to
\eqref{HP Hamiltonian} hold. The condition
$\bra{\psi(t)}b^\dagger b\ket{\psi(t)}\ll M$ sets a maximal time beyond which
\eqref{eq:SR_su(11)_comp_regimes} ceases to be a valid expression for the
symmetry-resolved Krylov complexity of the $\alpha_-$ mode. This time grows with
the number $M$ of atoms.

 A few limitations of this analysis should be stressed. First,
\eqref{eq:SR_su(11)_comp} is not the full complexity, but is restricted to the
$\mathfrak{su}(1,1)$ subsectors. As mentioned earlier, an analytical expression for the full
Krylov complexity lies beyond the current state of the art; the
symmetry-resolved analysis above, however, still provides an analytical handle on the
dynamics. On the other hand, away from resonance the Hamiltonian does not decompose into two $\mathfrak{su}(1,1)$ sectors any more and this inhibits an analytic derivation of the Krylov space and related complexity. It is worth stating that there still exists an underlying symmetry structure that is described by the Schr\"odinger or Jacobi algebra, though since it is of higher rank there is no clear prescription for the analytical distillation of the dynamics in Krylov space from it. As a matter of fact, previous works on the topic \cite{Haque:2022ncl,Patramanis:2023cwz} have highlighted that an exact answer is for the time being out of reach. 

This limitation can be partially overcome by examining the dynamics from the FSL perspective, as we will discuss in detail below.

\subsection{Dynamics on the FSL}

\begin{figure}[ht]
    \centering
    
\includegraphics[width=0.9\linewidth]{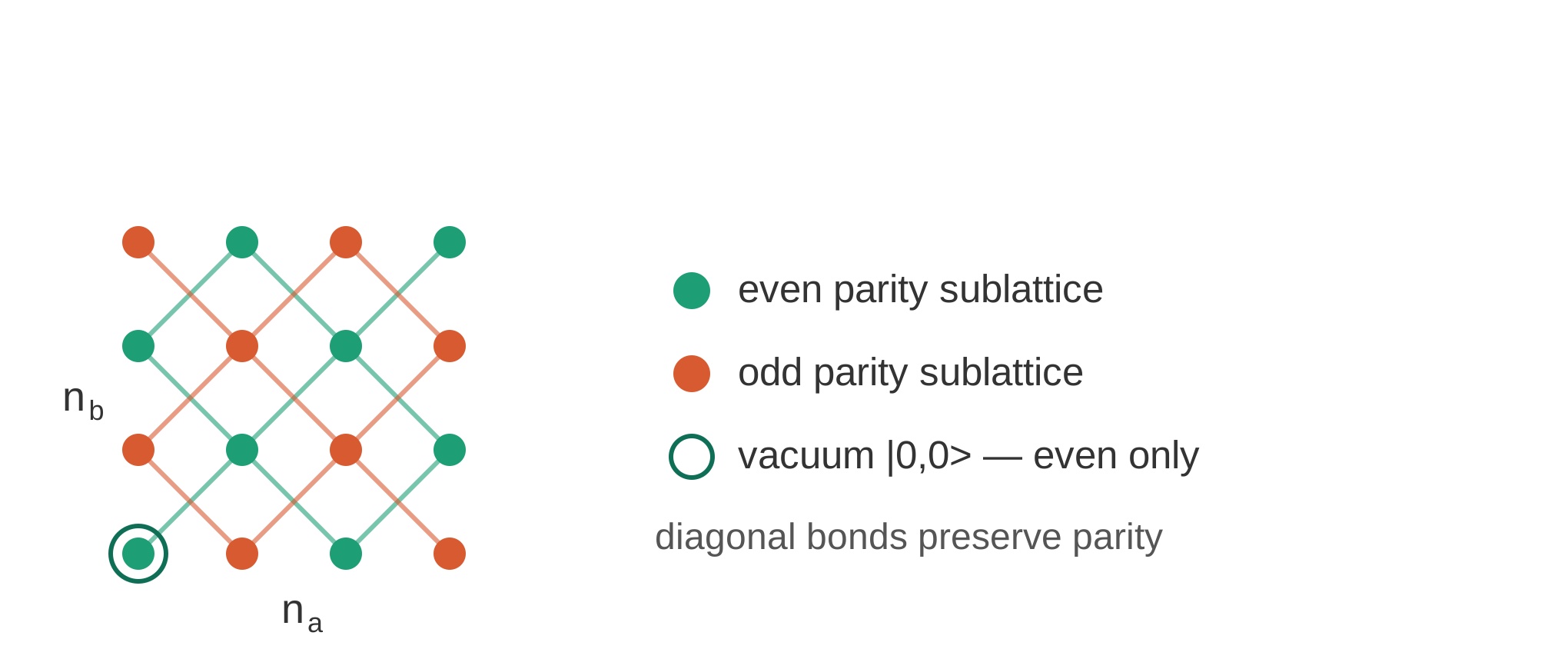}
    \caption{The FSL of the Dicke model \eqref{HP Hamiltonian} with vertices labeled by the occupation numbers $n_a$ and $n_b$, eigenvalues of $a^\dagger a $ and $b^\dagger b $ respectively. This lattice is decomposed in the two disconnected components (green and red) associated with the two parity subsectors  from the conservation of $(-1)^{a^\dagger a+b^\dagger b}$.
    If the initial state of the system is in either the even or odd parity sector, the Hamiltonian evolution keeps the system in the corresponding sublattice.}
    \label{bare Dicke Lattice}
\end{figure}

The FSL for the Dicke model \eqref{HP Hamiltonian} was already characterized in \cite{Saugmann:2022qzh} as an infinite square lattice, where every vertex is associated with a Fock state labeled by $n_a$ and $n_b$, eigenvalues of the occupation numbers $a^\dagger a $ and $b^\dagger b $ respectively. If we wish to be more precise it is actually composed of two superimposed, disconnected, square lattices, each associated with an eigenvalue of the parity operator $(-1)^{a^\dagger a+b^\dagger b}$ as we illustrate in Fig.\,\ref{bare Dicke Lattice}.
As mentioned in previous sections, 
quantifying spread over the FSL is an ambiguous endeavor. In this case, however, we can take advantage of the analysis of Sec.\,\ref{subsec:3mode_JC} and adapt the distances defined in that context to the square FSL of the Dicke model. In particular, if the initial state is $\ket{\psi_0}$ associated with the origin of the FSL, we can track the geodesic graph distance between the state evolving via \eqref{HP Hamiltonian} and $\ket{\psi_0}$, as well as the Euclidean distance.

\begin{figure}[ht]
    \centering
    \includegraphics[width=0.65\linewidth]{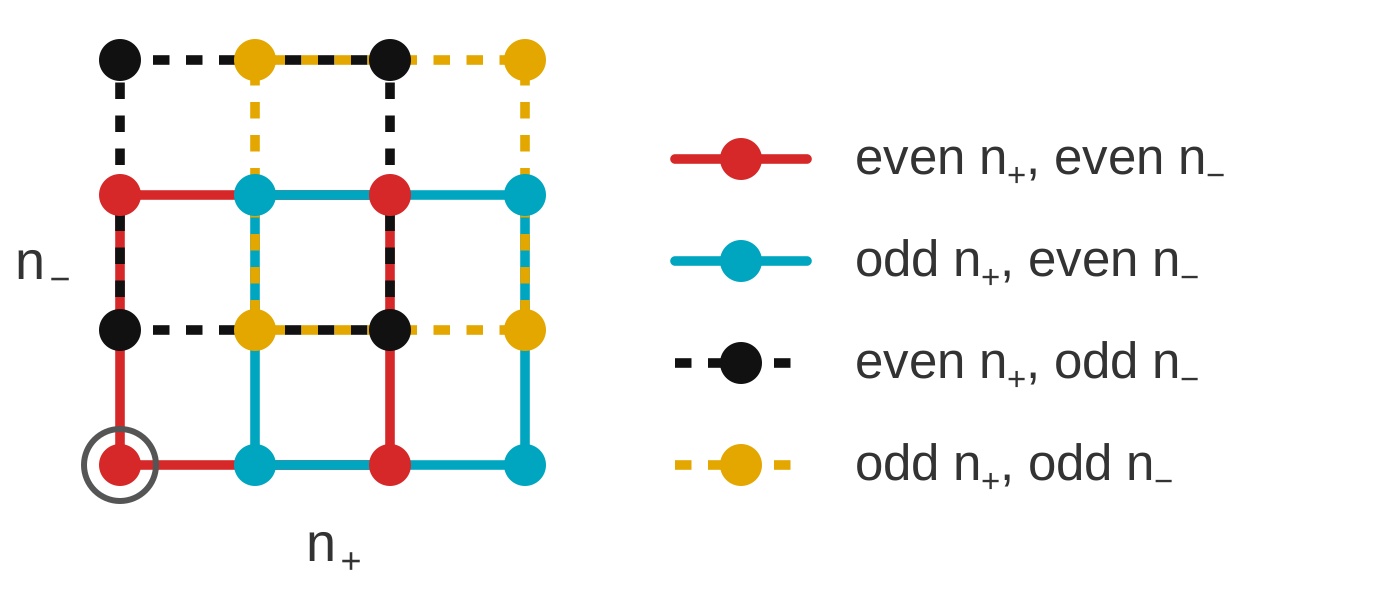}
    \caption{The FSL of the Dicke model at resonance, constructed from the Hamiltonian \eqref{eq: Hpolariton decomp}. The vertices are labeled by $n_+$ and $n_-$, eigenvalues of $c_+^\dagger c_+ $ and $c_-^\dagger c_- $ respectively. The four different colors identify four sublattices associated with the four parity sectors from the conservation of $(-1)^{c_+^\dagger c_+}$ and $(-1)^{c_-^\dagger c_-}$. The origin corresponds to the vacuum state $\ket{0_+}\ket{0_-}$ (bottom left in the figure).}
    \label{polariton sublattices}
\end{figure}

Let us again start from the resonant regime where we can directly compare with the Krylov picture we obtained above. In the polariton basis the FSL is still a square lattice with vertices labeled by the eigenvalues $n_+$ and $n_-$ of the number operators $c_+^\dagger c_+ $ and $c_-^\dagger c_- $ respectively. Due to the conservation of the parities $(-1)^{c_+^\dagger c_+}$ and $(-1)^{c_-^\dagger c_-}$ separately, the FSL further decomposes into sublattices as seen in Fig.\,\ref{polariton sublattices} where we have also rotated the lattice to distinguish between the original Fock and the polariton bases. 
If we then choose an initial state on a single sublattice, we find that the dynamics across the two spanning directions are independent, each described by a copy of $\mathfrak{su}(1,1)$. We illustrate that decomposition in Fig.\,\ref{polariton sublattices}.

Let us assume for simplicity that the system is initialized in the polariton vacuum state $\ket{0_+}\ket{0_-}$, corresponding to the origin of the FSL. Then the key observation is that the dynamics consists of motion along the spanning directions of the lattice. Tracking the average positions \\$\bra{\psi(t)}c^\dagger_+c_+\ket{\psi(t)}=\frac{g^2}{\alpha_+^2}\sin^2{(\alpha_+ t)}$ and  $\bra{\psi(t)}c^\dagger_-c_-\ket{\psi(t)}=\frac{g^2}{\alpha_-^2}\sin^2{(\alpha_- t)}$ along each direction is then equivalent to knowing the average position in the full lattice.
These two expectation values can be interpreted as coordinates on the
square FSL, and their time evolution traces out a trajectory in the
two-dimensional space in which the FSL is embedded, in close analogy with the
construction of Sec.\,\ref{subsec:Euclidean}.

\begin{figure}[ht]
    \centering
    \begin{subfigure}[b]{0.32\textwidth}
        \centering
        \includegraphics[width=\textwidth]{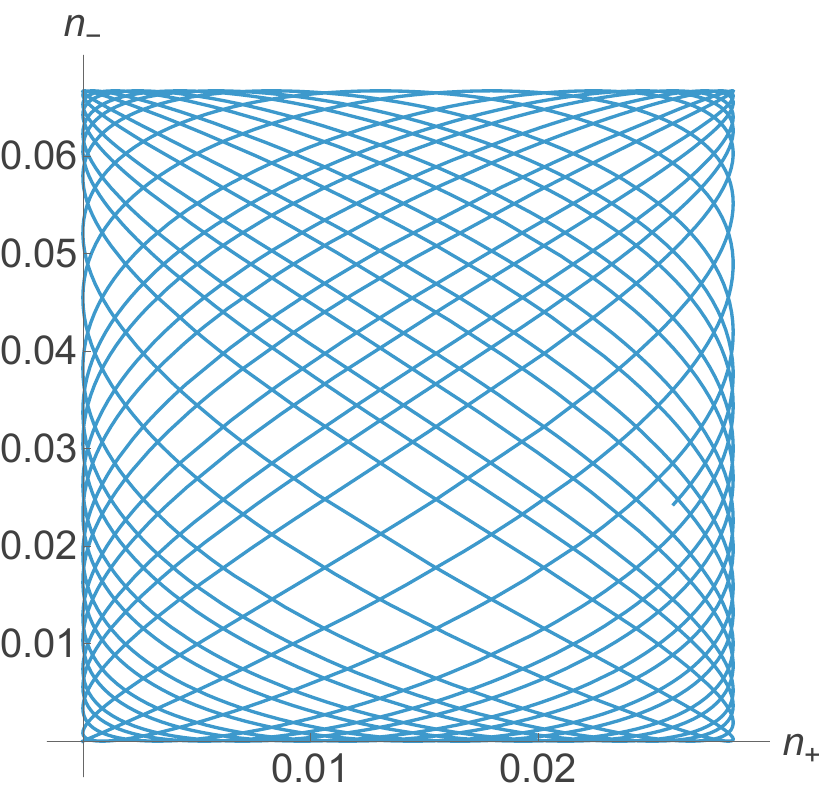}
        \caption{Trajectory on the FSL for $g=0.4 g_c$}
        \label{fig:sub1}
    \end{subfigure}
    \hfill
    \begin{subfigure}[b]{0.32\textwidth}
        \centering
        \includegraphics[width=\textwidth]{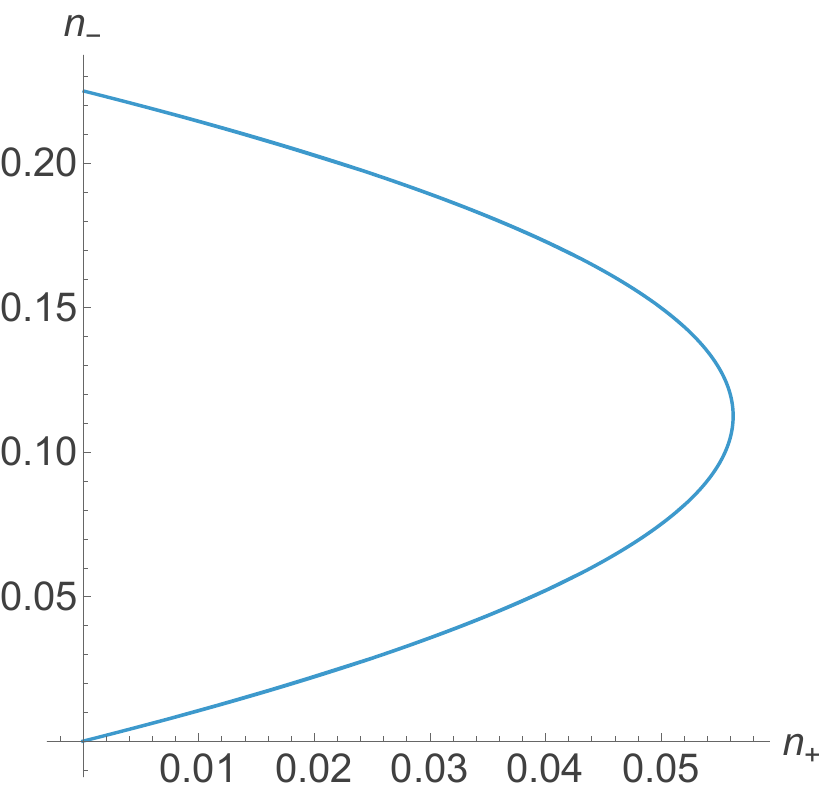}
        \caption{Trajectory on the FSL for $g=0.6 g_c$}
        \label{fig:sub2}
    \end{subfigure}
    \hfill
    \begin{subfigure}[b]{0.32\textwidth}
        \centering
        \includegraphics[width=\textwidth]{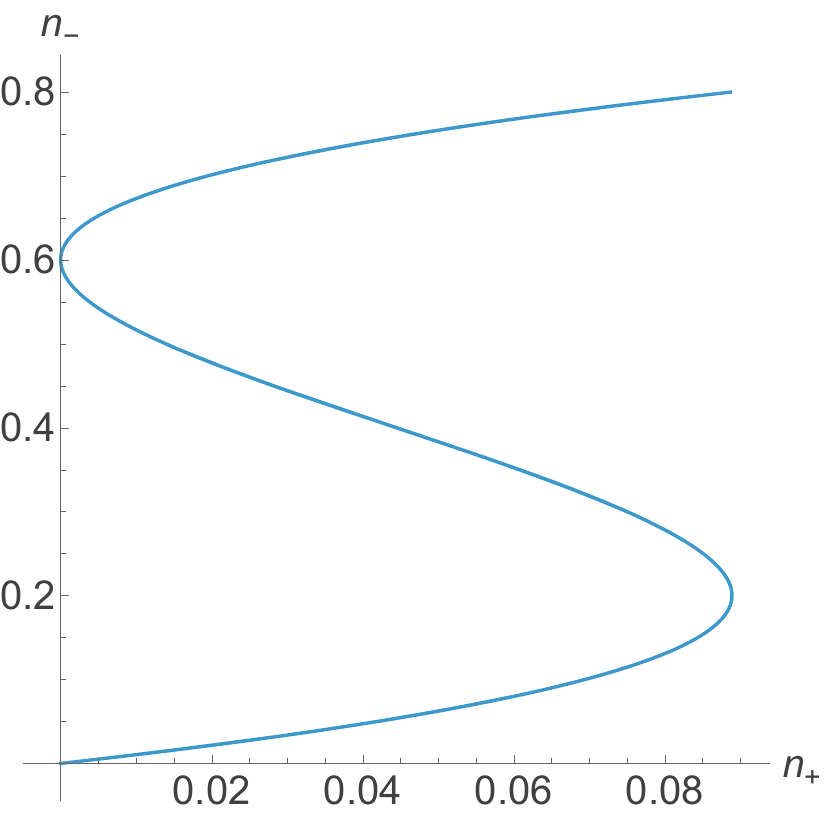}
        \caption{Trajectory on the FSL for $g=0.8 g_c$}
        \label{fig:sub3}
    \end{subfigure}
    \caption{Trajectories on the FSL with coordinates $\bra{\psi(t)}c^\dagger_+c_+\ket{\psi(t)}$ and $\bra{\psi(t)}c^\dagger_-c_-\ket{\psi(t)}$ for three different values of the coupling below the critical point. Note that while the first image depicts a curve that does not close on itself, leading to a volume-filling behavior, the other two lead to simplified dynamics.}
    \label{bounded curves}
\end{figure}

In the normal phase the trajectories on the FSL manifest as the collective motion of two harmonic oscillators, leading to Lissajous curves.  Generically these trajectories are volume-filling, but for special values of the coupling they only explore a limited portion of the space of states.
This happens when the ratio $\alpha_+/\alpha_-=\sqrt{(\omega+2g)/(\omega-2g)}$ is rational.
In both cases the trajectories are bounded, even though the FSL itself is infinite. In Fig.\,\ref{bounded curves} we show some examples illustrating the behavior that the system can assume in this regime. 

\begin{figure}[ht]
    \centering
    \begin{subfigure}[b]{0.45\textwidth}
        \centering
        \includegraphics[width=0.65\textwidth]{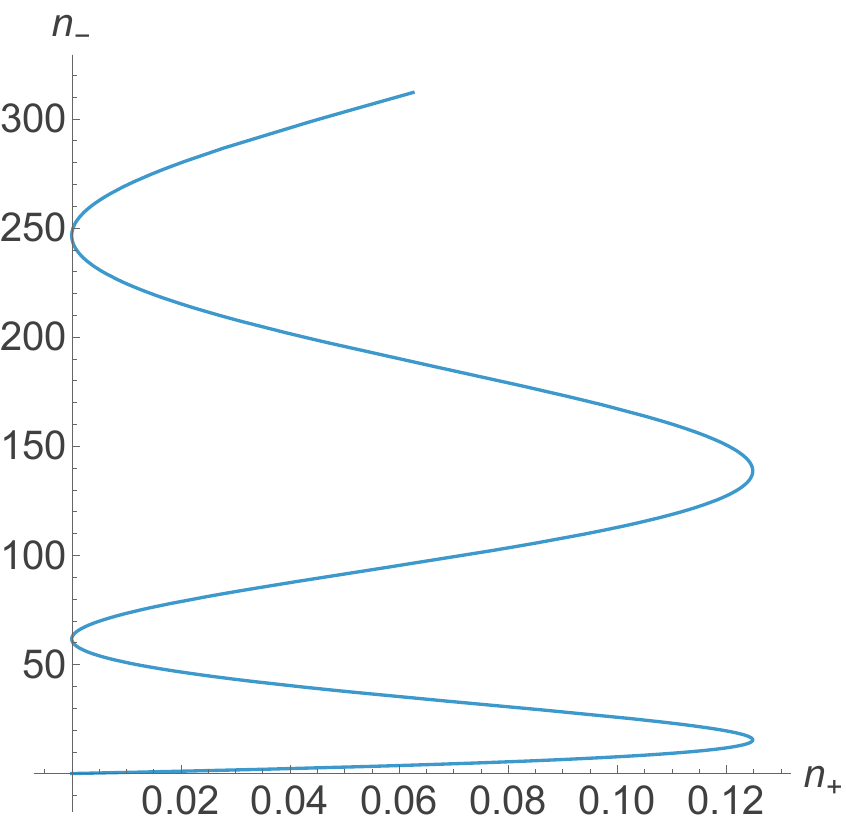}
        \caption{Trajectory on the FSL for $g\simeq g_c$}
        \label{fig:subb1}
    \end{subfigure}
    \hfill
    \begin{subfigure}[b]{0.45\textwidth}
        \centering
        \includegraphics[width=0.65\textwidth]{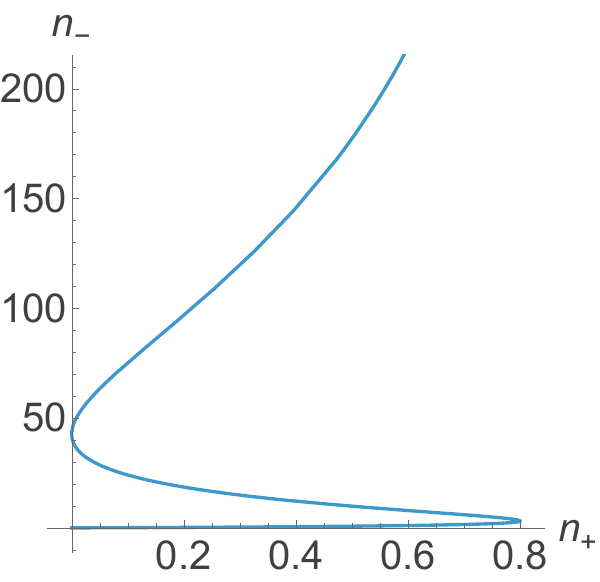}
        \caption{Trajectory on the FSL for $g=1.02 g_c$}
        \label{fig:subb2}
    \end{subfigure}
    \caption{Trajectories on the FSL with coordinates $\bra{\psi(t)}c^\dagger_+c_+\ket{\psi(t)}$ and $\bra{\psi(t)}c^\dagger_-c_-\ket{\psi(t)}$ for values of the coupling close to and above criticality. Note that in this case the trajectories become unbounded and deviate with quadratic or exponential rate respectively. 
    }
    \label{unbounded curves}
\end{figure}

On the other hand, close to the phase transition and in the superradiant phase the trajectories become unbounded and the average distance on the $n_-$ axis grows quadratically at the critical point and exponentially above it, as seen in Fig.\,\ref{unbounded curves}. 
This should not come as a surprise since in the resonant regime the information contained in the FSL is identical to what we extracted using the symmetry-resolved Krylov complexity \eqref{eq:SR_su(11)_comp} and \eqref{eq:SR_su(11)_comp_regimes}. 

Since the Dicke FSL is a simple square lattice structure, defining distances is also a much simpler endeavor compared to the JC model discussed in Sec.\,\ref{subsec:3mode_JC}. 
With this in mind, the first notion we can define is the graph distance, adapting \eqref{eq:graphdistance} and \eqref{eq:graphdistance_2ndexample} to the square FSL in Fig.\,\ref{polariton sublattices}. In this example, it is the natural quantum generalization of the Manhattan distance, and is given by the expression
\be
\label{eq:graphdista_rect}
d_\mathrm{G}(t)= \frac{1}{2}\bra{\psi(t)}c^\dagger_+c_+\ket{\psi(t)}+\frac{1}{2}\bra{\psi(t)}c^\dagger_-c_-\ket{\psi(t)}~.
\ee
Following the paradigm we set in the case of the multi-mode JC model and in particular equations \eqref{eq:coordinateN3n1} and \eqref{eq:coordinateN3hex}, we have included the factor of $\frac{1}{2}$ to rescale the distance between adjacent vertices on each of the four FSL components in Fig.\,\ref{polariton sublattices}. 

 In the case of resonance this expression builds a bridge between this FSL graph distance and the Krylov complexities  \eqref{eq:SR_su(11)_comp} since it is  precisely the sum of the symmetry-resolved Krylov complexities of the two modes. Alternatively, paralleling the definition in Sec.\,\ref{subsec:Euclidean}, we can define the Euclidean distance in terms of the two time-evolving coordinates introduced above
\be
\label{eq:euclid_dist_rect_avg}
d_\mathrm{E}(t)= \frac12\sqrt{\bra{\psi(t)}c^\dagger_+c_+\ket{\psi(t)}^2 +\bra{\psi(t)}c^\dagger_-c_-\ket{\psi(t)}^2}~,
\ee
which corresponds to \eqref{eq:Euclideandist} adapted to the present example.
Since the average and the square root do not commute, we can also consider
\be\label{eq:euclid_dist_rect}
\tilde{d}_\mathrm{E}(t)= \frac12\bra{\psi(t)}\sqrt{(c^\dagger_+c_+)^2 +(c^\dagger_-c_-)^2}\ket{\psi(t)}~,
\ee
and compare it with \eqref{eq:euclid_dist_rect_avg}. 
Let us remark that all the measures
\eqref{eq:graphdista_rect}-\eqref{eq:euclid_dist_rect} are reliable distances
on the FSL of the Dicke model only within the time window set by the condition
$\bra{\psi(t)}b^\dagger b\ket{\psi(t)}\ll M$ discussed above.

\begin{figure}[ht]
    \centering
    \begin{subfigure}[b]{0.32\textwidth}
        \centering
        \includegraphics[width=\textwidth]{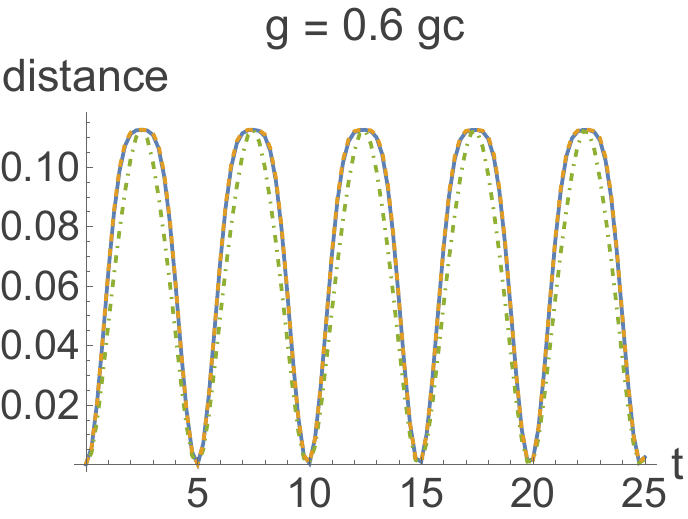}
        \label{pl}
    \end{subfigure}
    \hfill
    \begin{subfigure}[b]{0.32\textwidth}
        \centering
        \includegraphics[width=\textwidth]{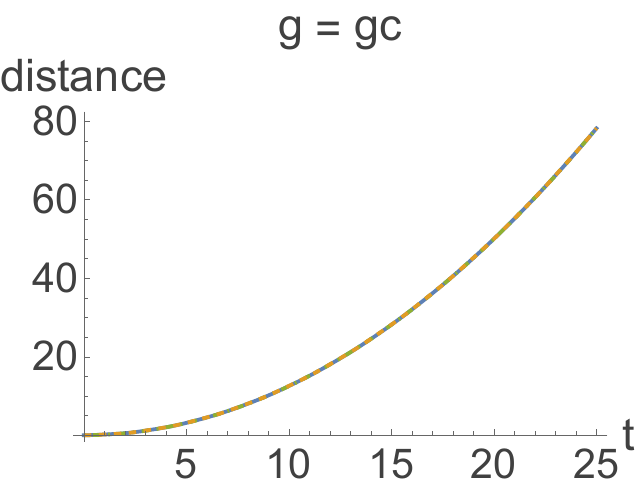}
        \label{pla}
    \end{subfigure}
    \hfill
    \begin{subfigure}[b]{0.32\textwidth}
        \centering
        \includegraphics[width=\textwidth]{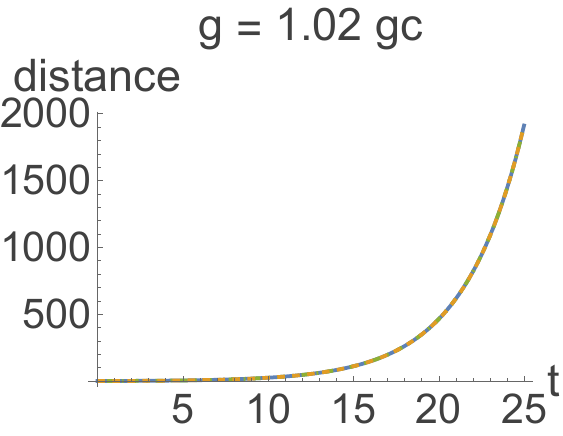}
        \label{plac}
    \end{subfigure}
    \caption{Comparison of different measures of distance as functions of time. The yellow dashed line corresponds to the graph distance $d_{\rm G}$ in \eqref{eq:graphdista_rect}, the green dashed line corresponds to $d_{\rm E}$ in \eqref{eq:euclid_dist_rect_avg}, and the blue line corresponds to the distance $\tilde d_{\rm E}$ in \eqref{eq:euclid_dist_rect}.}
    \label{on-resonance comparison}
\end{figure}

 Although in the resonant regime the problem simplifies and the joint distribution of the two polariton modes factorizes into two marginals, there is no closed form that the  distance $\tilde{d}_\mathrm{E}$ admits, leading us in turn to compute it numerically, expanding \eqref{eq:euclid_dist_rect} in the basis of occupation numbers. A comparison among these three measures of distance can be found in Fig.\,\ref{on-resonance comparison}, where one can easily verify that in this example all the quantities provide a very similar description of the system dynamics. In particular, all three of the proposed FSL distances detect the quantum phase
transition: their late-time growth changes from bounded oscillations to
quadratic and then exponential growth as the critical coupling is crossed.

Let us now move away from resonance, where a decomposition in terms of separate symmetry sectors no longer exists. As argued previously this is detrimental for the calculation of Krylov complexity and related quantities, but probing the dynamics of the system is still feasible through the FSL. Using the off-resonance expression for $\alpha_\pm$ 
\be
\alpha_\pm^2=\frac{\omega^2 +\Delta^2}{2}\pm \frac{1}{2} \sqrt{(\omega^2-\Delta^2)^2+16g^2\omega \Delta}~,
\ee
and writing $\alpha_+^2-\alpha_-^2=\Xi$ for the sake of brevity, the graph distance \eqref{eq:graphdista_rect} can be written as
\be
\label{eq:offi diag dist}
\begin{split}
    d_\mathrm{G}(t)=&\frac{2g^4(\omega +\Delta)^2}{\Xi^2}\left(\frac{\sin^2{(\alpha_+ t)}}
    {\alpha_+^2}+\frac{\sin^2{(\alpha_- t)}}{\alpha_-^2}\right)+\\
    &\frac{2g^2(\omega -\Delta)^2}{\Xi^2}\left(1-\cos{(\alpha_+t)} \cos{(\alpha_-t)}+\frac{\omega\Delta -2g^2}  {\alpha_+\alpha_-}\sin{(\alpha_+t)} \sin{(\alpha_-t)}\right)~.
\end{split}
\ee
The derivation of this formula is reported in Appendix \ref{app:off-resonance}.
Strictly speaking, \eqref{eq:graphdista_rect} counts steps on a graph only at
resonance, where the polariton lattice is square, as shown in Fig.\,\ref{polariton sublattices}, and the dynamics proceeds
along its two spanning directions. Away from resonance the term proportional to
$\omega-\Delta$ in \eqref{eq:HPHamiltonian_polar} connects the polariton occupation number eigenvalues $(n_+,n_-)$ to
$(n_+\pm1,n_-\mp1)$, and that counting no longer applies. 
Nonetheless \eqref{eq:offi diag dist} remains an insightful measure of the spread of the state over the FSL, and we shall keep referring to it as a graph distance owing to the fact that it is a direct generalization of the graph distance at resonance. For a more rigorous analysis of this point one needs to modify the definition of distance as explained in Appendix \ref{app:Wasserstein}. 

For the sake of clarity, we remark that whenever $\alpha_-$ becomes purely imaginary, i.e.\ when the system enters the superradiant phase, \eqref{eq:offi diag dist} is to be understood as its analytic continuation in $\alpha_-$, as already done in the last of eqs.~\eqref{eq:SR_su(11)_comp_regimes}. The substitution $\alpha_-=i|\alpha_-|$ turns the oscillating functions of $\alpha_- t$ into hyperbolic ones, and the result stays real. The change of $\alpha_-$ from real to purely imaginary occurs at
$g=g_c=\sqrt{\omega\Delta}/2$, i.e. at the transition to the superradiant
phase, now for generic values of $\omega$ and $\Delta$. This transition is
detected by the late-time scaling of \eqref{eq:offi diag dist}.
Also for this last formula, it is important to stress that it provides a reliable prediction for the Dicke model in the regime $M\gg 1$ and around the phase transition and in the superradiant phase as long as the values of time is not too large to violate the condition $\langle \psi(t) |b^\dagger b|\psi(t)\rangle \ll M$.

\begin{figure}[ht]
    \centering
    \includegraphics[trim={50 0 0 0}, clip, width=\linewidth]{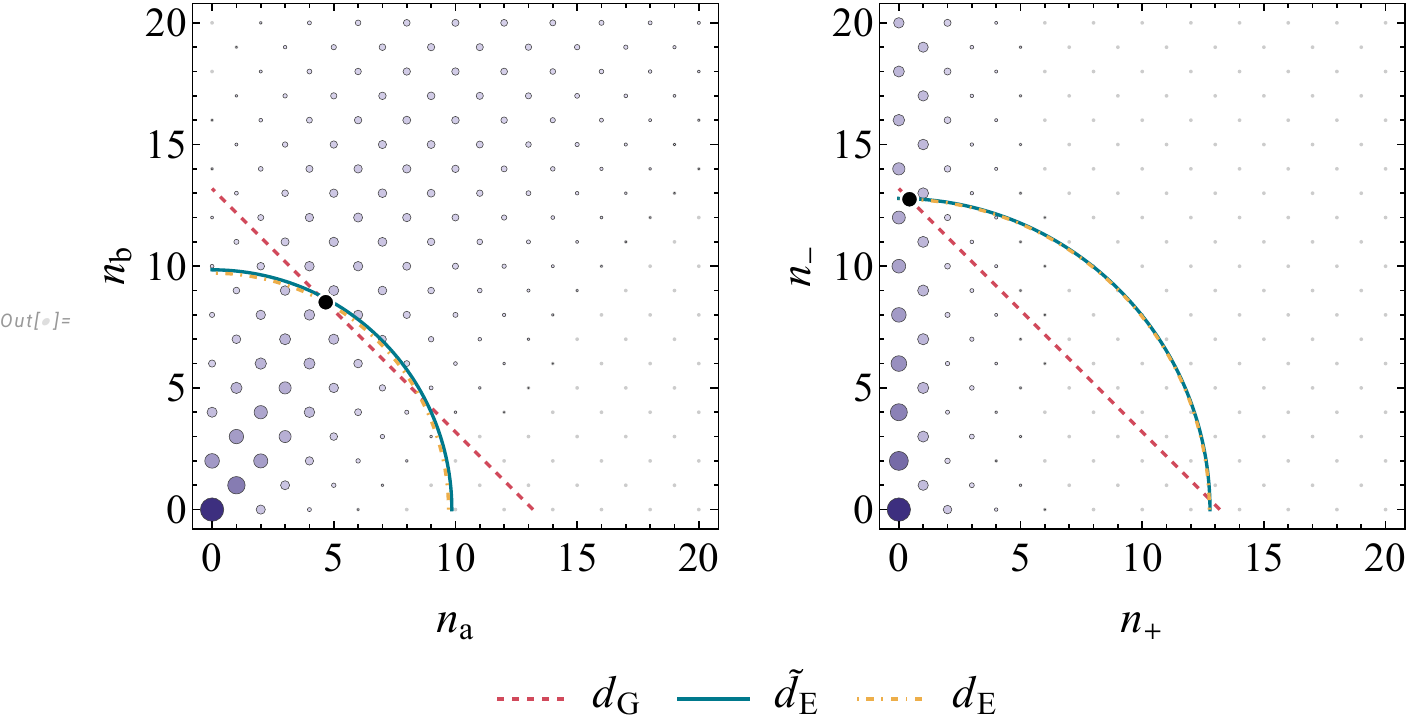}
    \caption{Different measures of distance on the FSL of the off-resonance Dicke model in the thermodynamic limit, plotted for $g\simeq g_c$. The size and hue of the vertices correspond to the probability distribution over the lattice: a larger vertex indicates a higher probability associated with that Fock state. The red dashed line in both panels denotes the graph distance $d_{\rm G} $  in \eqref{eq:graphdista_rect}. In the right panel, the yellow dashed curve is the distance $d_{\rm E} $ in \eqref{eq:euclid_dist_rect_avg} and the green curve the distance $\tilde d_{\rm E} $ in \eqref{eq:euclid_dist_rect}. In the left panel \eqref{eq:euclid_dist_rect_avg} and \eqref{eq:euclid_dist_rect} have been modified by replacing $c_+\to a$ and $c_-\to b$ and are shown as yellow dashed and green solid curves respectively. The black dot represents the average occupation number. The left panel shows the FSL of the bare Hamiltonian \eqref{HP Hamiltonian} whereas the right one shows the FSL from the Hamiltonian \eqref{eq:HPHamiltonian_polar} in the polariton basis. Different values of the coupling lead to similar pictures though for the normal phase there is strong localization near the initial state and in the superradiant phase the delocalization present above is more pronounced.
    }
    \label{off resonance distances}
\end{figure}

While the expression in \eqref{eq:offi diag dist} is much less palatable than what we obtained for the resonant regime, it exactly describes the dynamics of the model for any values of the parameters in the Hamiltonian. One can also readily check that for $\omega=\Delta$ the above expression reduces to the resonant case.
For a complete analysis, we can generalize also \eqref{eq:euclid_dist_rect_avg} and \eqref{eq:euclid_dist_rect} to the off-resonance case or, alternatively, we can replace $c_+\to a $ and $ c_-\to b $, and compute the resulting quantities for general values of the parameters. Since there is no closed form expression for $\tilde d_{\textrm{E}}$, we perform a direct graphical comparison in Fig.\,\ref{off resonance distances}.
A prominent feature of Fig.\,\ref{off resonance distances} is that while we can obtain an accurate picture of the average distance, the measures under consideration fail to capture the variance of the distribution, or in other words the spreading over the lattice. This effect is captured by the relative sizes of the vertices which are also a basis-dependent observable. This is similar to what we encountered in Sec.\,\ref{multi-mode} where we computed the participation and K-entropies as probes of this phenomenon. Here the relevant object for our analysis is the phase transition and, since that is already visible using the diagnostics presented in Fig.\,\ref{off resonance distances}, we restrict ourselves to  this qualitative comment rather than a complete investigation of these characteristics of the distribution.

\section{Conclusions} \label{conclusions}

In this work we have taken a step forward in representing generic unitary
quantum dynamics as an evolution on a lattice. In order to go beyond the one-dimensional
reduction inherent in the Krylov space approach, where the dynamics is always
projected on a chain, we have studied quantum evolutions as represented on the
FSL of the underlying system, which is in general a higher-dimensional lattice.
Apart from the dimensionality of the resulting lattices, the two methods differ
in another important respect. While the FSL components depend only on the
evolution Hamiltonian and its symmetries, the Krylov basis, and hence the
Krylov chain, depends strongly on the initial state as well.

After a review of the two approaches and a preliminary comparison in
Sec.\,\ref{review section}, we have applied them to various prototypical
quantum optics models: the Rabi and JC models, together with their
multi-atom and multi-mode generalizations respectively. For the single-mode models in Sec.\,\ref{intro to qom} and for the two-mode JC model in Sec.\,\ref{subsec:twomode_JC}, we observe that the FSLs are disconnected chains which can be treated as Krylov chains as long as the initial states fall within the corresponding subsector and represent one of the extremal vertices of the FSL itself. In that sense, whether we view the dynamics through the lens of the Fock or Krylov space is just a matter of semantics, as these frameworks provide the same information about the system.
Notably, we can confirm the non-chaotic behavior of the Rabi model through the computation of the Lanczos coefficients and of the Krylov complexity. Whether the model is really quantum integrable remains debatable (see \cite{Batchelor2015} as a counterpoint to \cite{Braak:2011tmj}) and while our results are not capable of resolving this issue, they constitute a robust indication that the Rabi model is in a different class from fast scrambling and chaotic systems.

For the JC model with $N\geq3$ photonic modes, and for the Rabi model with more
than one two-level atom, also known as the Dicke model, the situation is
different since the FSL is a higher-dimensional lattice. In
Sec.\,\ref{subsec:3mode_JC} we have analyzed the simplest of these cases, the
three-mode JC model, both in the large-detuning limit and away from it. The
Krylov space analysis can still be carried out, at least numerically, and the
corresponding Krylov complexity computed. In contrast, the higher dimensionality of the FSL makes the definition of a distance between the initial and the
evolving state less natural. We have provided two tools to study the dynamics
from this perspective: the graph distances \eqref{eq:graphdistance} and
\eqref{eq:graphdistance_2ndexample} for the two regimes, and the Euclidean distance
\eqref{eq:Euclideandist}, evaluated with the coordinates
\eqref{eq:coordinateN3n1} or \eqref{eq:coordinateN3hex} according to whether the
trajectory is tracked in the ambient space of the FSL in or away from the large
detuning regime. The numerical analysis reported in
Fig.\,\ref{fig:timeaverages} shows that these two distances carry very similar
information and behave much like the Krylov complexity. This makes them
reliable probes of the size of the region of the Hilbert space explored by the
dynamics. Fig.\,\ref{fig:spread}, on the other hand,
shows that the spread measured by the participation entropy in the Fock
basis and by the K-entropy in the Krylov basis behaves differently in the two
cases, thus confirming that the two bases act as different lenses on the same
quantum dynamics.

The analysis in Sec.\,\ref{multi-mode} highlights some difficulties in defining
distances and spread measures on FSLs, due to their strong dependence on the
model. This is in contrast with the Krylov space approach, which, by always
projecting the dynamics onto a one-dimensional chain, provides natural measures
such as the Krylov complexity. At the same time, our findings support the
usefulness of studying a dynamics through different bases and their associated
lattice models, and this is particularly relevant when one aims at analytical
insight. Indeed, the Krylov data can be determined in closed form only in a few
instances, and the task becomes especially hard in many-body systems, whereas
the Fock basis is typically more readily available and better understood.
Studying the dynamics from the FSL perspective can therefore yield analytical
results in cases where the Krylov complexity cannot be computed exactly.

This is what happens in the example of Sec.\,\ref{Dicke}, where we have studied
the Dicke model in the thermodynamic limit from the viewpoint of both the (symmetry-resolved) Krylov space
and the FSL. In the resonant regime the Hamiltonian splits into two $\mathfrak{su}(1,1)$ sectors, and each sublattice of the FSL factorizes into the product of the two
corresponding Krylov chains. Thus, while the full Krylov chain cannot be derived
analytically, its symmetry-resolved counterpart can, and the tools developed in
Sec.\,\ref{multi-mode} carry over to this FSL. Away from resonance the
decomposition is lost and the symmetry-resolved Krylov complexity is no longer accessible, yet the
graph distance is still available in closed form, see
\eqref{eq:offi diag dist}. The main finding is that the quantum phase transition
to the superradiant phase is detected by all these quantities: the graph
distance, the two Euclidean distances adapted to this FSL and, in the resonant regime, the
symmetry-resolved Krylov complexity all change their scaling in time across the
critical coupling, passing from bounded oscillations to quadratic and then
exponential growth. They therefore provide further proxies for quantum phase
transitions, built on spread measures of the quantum evolution.
All the time-dependent analytical results of Sec.\,\ref{Dicke} are reliable as
long as $\bra{\psi(t)}b^\dagger b\ket{\psi(t)}\ll M$, the condition under which
\eqref{HP Hamiltonian} provides an effective description of the Dicke model. This is
automatically satisfied at all times in the normal phase, and defining a
maximal time at the critical point and above it. It is important to emphasize that the limitation imposed by the thermodynamic limit and the way it affects our description of the superradiant phase is compatible with a realistic manifestation of the system. Namely, the natural description of the Dicke model ground state in the superradiant phase involves a macroscopic number of Fock states and therefore the perspective we provided by tracking the evolution of the lowest weight state under the Dicke Hamiltonian has to inevitably break down. Nevertheless, our aim was the detection of the phase transition which is feasible under the conditions we have set. The study of the rescaled eigenstates of the Hamiltonian above the phase transition is a separate problem going beyond the scope of this work, which is however of great interest as it could potentially probe the genuine chaotic aspects of this system that our analysis is not sensitive to.

This work opens the way to a broad program with the goal of understanding quantum dynamics by
representing it as an equivalent model evolving on a lattice. This program can
be extended in several directions, beyond the Krylov and the Fock spaces. Other
choices of basis would provide further complementary perspectives, each with
its own lattice and its own natural notions of distance and spread. In
particular, the quest for a precise classification of dynamics ranging from
integrable to chaotic may benefit from the fine-grained information that can be
obtained by analyzing a quantum evolution from many different angles, that is,
in different bases and lattice representations.
Analyses along these lines could also provide useful proxies for distinguishing
quantum chaos from saddle-dominated scrambling \cite{Xu:2019lhc,Bhattacharjee:2022vlt,Huh:2023jxt,Camargo:2026szl}. Since the Dicke model studied in the present manuscript has been shown to exhibit saddle-dominated scrambling \cite{Pilatowsky-Cameo:2019qxt}, this model can be an insightful playground to address this type of questions.
In addition, since the Fock basis is given from the definition of the model and does not require
a Gram-Schmidt construction, the FSL could be an interesting platform for
devising spread and complexity measures for time-dependent Hamiltonians and for
open quantum systems \cite{Naves:2025iec}. In this respect, a variation of the JC Hamiltonian \eqref{eq:JCham_1mode} with a
time-dependent coupling $g(t)$ can serve as a useful playground. In this case
the FSL still consists of an infinite set of decoupled one-dimensional two-site chains. If a further photon drive term $\propto a+a^\dagger$ is
added, these chains become coupled and the FSL turns two-dimensional \cite{Larson:2024rmo}. This
is an appealing setup for studying dynamics generated by time-dependent
Hamiltonians in a case where the FSL and the Krylov chain have different
dimensionalities.

Our analysis may also serve as a bridge, bringing tools from various branches of
mathematics into the study of quantum dynamics. Describing quantum evolution as 
lattice dynamics opens interesting connections with graph theory and with
quantum random walks, which have already been related to Krylov and FSL methods
in \cite{Patramanis:2026jtp,Ferraro:2026rrq}. In this context, much remains to be explored, such as the use of other graph distances, including the Hamming distance, which would potentially allow us to get closer to the language of the quantum information community. Moreover, since the evolution
induces a time-dependent probability distribution over the lattice, we envisage
the use of tools from optimal transport, as discussed preliminarily in
\cite{Hashimoto:2023swv,Hashimoto:2026kjy,Hashimoto:2026zlh} and in
Appendix~\ref{app:Wasserstein} of this manuscript.

As detailed throughout the manuscript, the structure of the FSL 
can be inferred
from the Hamiltonian, same as the Krylov chain once the
evolution Hamiltonian is projected onto the Krylov basis. The Krylov chain,
however, enjoys a further property: it can be fully reconstructed from a single
function of time, the return amplitude.
It would be interesting to understand whether the properties of the FSL can
likewise be extracted from a single suitable function\footnote{We thank Paweł Caputa for raising this question.}.

The graph distance introduced in \eqref{eq:graphdistance}, and extended to other
models in \eqref{eq:graphdistance_2ndexample} and \eqref{eq:graphdista_rect},
rests on a minimization: it is the quantum generalization of the minimal number
of steps needed to connect two sites of the FSL. This minimization is implicit
in the expressions above and becomes explicit in the optimal-transport
formulation of Appendix~\ref{app:Wasserstein}. Defining complexity measures as
minimal distances on a space of states is precisely the logic underlying Nielsen
complexity \cite{Nielsen:2006cea,Jefferson:2017sdb}.
While the relationship between Krylov and Nielsen complexities has been the subject of previous works \cite{Craps:2023ivc,Craps:2025kub}, a fully geometric understanding of the matter is still lacking. Given the natural geometric properties (and ambiguities) of the FSL framework, it can be viewed as a more natural candidate for bridging these two subjects, potentially providing the missing link between Nielsen and Krylov approaches. 
 
To go beyond the Krylov chain description while retaining full analytical
control over the dynamics, one can consider Hamiltonians with dynamical symmetry generated by a Lie algebra 
and a highest (or lowest) weight initial state. If the algebra has only one set of ladder operators, Krylov and Fock spaces
are intimately related, as shown in Appendix~\ref{Krylov-Fock equivalence}, while
for more general algebras this is no longer true. In those cases, the FSL is a promising
tool to represent these dynamics \cite{Ferraro:2026twt}, and we regard the
generalization of Krylov complexity and of other spread measures to this
framework as an intriguing research direction \cite{DiGiulio:WIP}.

More generally, other approaches that describe many-body quantum systems and
their dynamics in terms of simpler lattice models can also be framed within our
program. For instance, the commutativity graph method developed in
\cite{Wang:2019jsj} can be thought of as an operator-based generalization of the
FSL discussed in the present manuscript. The method of
\cite{Chin:2010erz,Prior:2010nnl}, instead, makes it possible to map a class of
models onto Hamiltonians with nearest-neighbor hopping terms only, resembling
the outcome of the Gram-Schmidt procedure in the Krylov-space approach.
Conversely, one can reverse the logic and reproduce the dynamics of many-body
systems by constructing higher-dimensional lattices whose couplings depend on
position in the same way as the Lanczos coefficients that make the Krylov
dynamics analytically solvable. 

Last but not least, our work constitutes an important development in connecting theoretical considerations to tangible physical systems and practical applications. The models we considered, apart from having a long history of experimental realizations \cite{Larson:2021flt, Forn-Diaz:2019epp, FriskKockum:2018nuz}, offer concrete manifestations of some of the universal symmetry classes first described in \cite{Caputa:2021sib,Balasubramanian:2022tpr}. It is worth emphasizing in particular the results of Sec.\,\ref{Dicke}, in which we find the Dicke model spanning all the possible realizations of the standard $\mathfrak{su}(1,1)$ results, mediated by the superradiant phase transition. As experiments that directly probe the dynamics of complex quantum systems become increasingly feasible, we have been able to measure quantities that have had a significant theoretical impact. This is especially prominent when it comes to the measurement of out-of-time-ordered correlators (see for example \cite{Li:2016xhw,Garttner:2016mqj,Landsman:2018jpm,Blok:2020may}). We hope that our results can serve in a similar capacity, providing a potential avenue for the measurement of the quantities we studied and upgrading our arsenal in the endeavor of characterizing and understanding these types of quantum systems that hold the key to unraveling a host of mysterious phenomena ranging from black holes to emergent properties of ordinary matter.

\section*{Acknowledgments}

We wish to thank Paweł Caputa, Andr{\'a}s Grabarits and Caio Naves for insightful discussions, and Paweł Caputa, and Sara Murciano for feedback and suggestions on the manuscript. We especially thank Jonas Larson, whose feedback have been instrumental in the presentation of the FSL and quantum optics portions of the manuscript.

Several of the figures that illustrate the different FSLs were generated with the help of Claude. We bear the responsibility for the accuracy and validity of the results presented therein. 

The work of GDG is supported by the ERC Consolidator grant (number: 101125449/acronym: QComplexity).  Views and opinions expressed are however those of the authors only and do not necessarily reflect those of the European Union or the European Research Council. Neither the European Union nor the granting authority can be held responsible for them.

The work of DP is supported by the National Science Foundation of China (NSFC) grant NO. 12447108.

\appendix

\section{Fock and Krylov bases for Lie algebras with one set of ladder operators} \label{Krylov-Fock equivalence}

In this appendix we show that Fock and Krylov bases are equivalent when constructed through Lie algebras with one set of ladder operators. As stated in the main text this result is implicit in previous works, but this is a good opportunity to pose it in its full generality. An important ingredient is the Lanczos uniqueness theorem (see \cite{viswanath1994recursion,Parlett}) which states that given a general operator $A$ and a basis of orthonormal vectors $Q=(q_1,q_2,...q_n)$, the tridiagonal form of $A$ with positive off-diagonal entries given by $Q^\dagger AQ=T$ is uniquely determined by $A$ itself and $q_1$. 

The additional structure we assume here is the existence of a Lie algebra with one set of ladder operators which implies the following Cartan-Weyl basis
\be
[C,E_\pm]= \pm\alpha E_\pm ,\quad [E_+,E_-]=f(C)~,
\ee
where $\alpha$ is a constant and $f$ is a polynomial. We also assume that this algebra furnishes a unitary irreducible representation on a Hilbert space containing a vector $\ket{0}$ such that $E_-\ket{0}=0$ and $C\ket{0}=\lambda \ket{0}$. The unitarity enforces $E^\dagger_+=E_-$ and $C^\dagger=C$.

The action of the generators in this representation assumes the form
\be
C E^n_+\ket{0}=(\lambda +\alpha n) \ket{n}, \quad E_-E_+\ket{n}=(E_+E_- -f(C))\ket{n}~.
\ee
By defining recursively the action 
\be
E_+\ket{n}=\sqrt{c_{n+1}}\ket{n+1},
\quad E_-\ket{n+1}=\sqrt{c_{n+1}}e^{i\phi_n}\ket{n}~,
\ee
we find that $c_{n+1}=c_n-f(\lambda+n\alpha)$. Without loss of generality, we can reabsorb the phases $e^{i\phi_n}$ into the definition of $\ket{n}$, setting them to one. Clearly, a dynamical Hamiltonian which is a linear combination of the algebra generators $H=a_+ E_+ +a_- E_- +a_0 C+ \text{cst}$ is going to have the following action on the Fock basis 
\be \label{eq:A4}
H\ket{n}=a_+\sqrt{c_{n+1}}\ket{n+1}+a_-\sqrt{c_n}\ket{n-1}+a_0(\lambda +\alpha n)\ket{n} +\text{cst}\ket{n}~,
\ee
where $a_+=a_-^*$ by Hermiticity,  $a_+\neq0$ and $c_n\geq0$, with the equality becoming valid when the recursion stops. This immediately leads to the identification 
\be
b_n=\abs{a_+}\sqrt{c_n},\quad a_n= a_0(\lambda+\alpha n) +\text{cst}~,
\ee
which along with the Lanczos uniqueness theorem concludes the proof that the Fock basis is identified with the Krylov basis. We remind the reader that crucially this is only true under the assumption that the initial state is a highest weight state \footnote{This is a slight abuse of terminology since we are technically using lowest weight states. However, one typically refers to any  extremal weight state as ``highest" so we believe our linguistic abuse is benign}.
Depending on the choice of initial state this identification will generically fail due to the sensitivity of the formation of the Krylov basis from different seed vectors. 

It is straightforward to obtain the familiar examples of Heisenberg--Weyl, $\mathfrak{su}(1,1)$ and $\mathfrak{su}(2)$ via the following identifications.
\begin{itemize}
\item \emph{Heisenberg--Weyl:} $E_+=a^\dagger$, $E_-=a$, $C=a^\dagger a$, so that $\alpha=1$, $f=-1$ and $\lambda=0$. Equation \eqref{eq:A4} gives $c_n=n$. For instance, $H=\omega a^\dagger a+g(a+a^\dagger)$ has $a_n=\omega n$ and $b_n=g\sqrt n$.
\item \emph{su(1,1):} $E_\pm=K_\pm$, $C=K_0$, so that $\alpha=1$ and $f(C)=-2C$. With Bargmann index $\lambda=k$, \eqref{eq:A4} gives $c_n=n(n+2k-1)$. For the polariton Hamiltonians \eqref{eq: Hpolariton decomp}, the vacuum has $k=1/4$ and $H_\pm=\pm g(K_++K_-)+2(\omega\pm g)K_0-(\omega\pm g)/2$. This gives $a_n=2(\omega\pm g)n$ and $b_n=g\sqrt{n(n-1/2)}$.
\item \emph{su(2):} $E_\pm=J_\pm$, $C=J_0$, so that $\alpha=1$ and $f(C)=2C$. The highest-weight state has $\lambda=-j$, and \eqref{eq:A4} gives $c_n=n(2j-n+1)$. This vanishes at $n=2j+1$, reproducing the $(2j+1)$-dimensional representation. In the two-mode JC model of Sec.\,\ref{subsec:twomode_JC} at large detuning, $J_+=a_2^\dagger a_1$ and $J_0=(a_2^\dagger a_2-a_1^\dagger a_1)/2$, with $2j=n_1+n_2$. 
\end{itemize}

\section{A Wasserstein distance on the Fock state lattice}
\label{app:Wasserstein}

In this appendix, we discuss a generalization of \eqref{eq:graphdistance} and \eqref{eq:graphdistance_2ndexample}, which allows us to quantify the graph distance between a generic initial
state in a fixed charge sector of the three-mode JC model and the state evolved
from it. This makes it possible to include in
the analysis of Sec.\,\ref{multi-mode} initial states that are not necessarily
represented by the corners of the FSL.  

In what follows, we focus mainly on the three-mode JC model in the large
detuning limit, whose FSL is the triangular graph in the right panel of
Fig.\,\ref{fig:JC_N3_FSLs}. Since all the states in a given FSL component have
the same value of the atomic excitation, each site is labeled by the three bosonic
occupation numbers only, which we collect in a single vector
$\boldsymbol{n}=(n_1,n_2,n_3)$.

To construct a general quantum version of the graph distance, we start from the
classical graph distance defined on the FSL graph. Given two vertices
$\boldsymbol{n}$ and $\boldsymbol{m}$, their classical graph distance
$d(\boldsymbol{n},\boldsymbol{m})$ is the minimal number of steps on the graph
connecting them,
\begin{equation}
d(\boldsymbol{n},\boldsymbol{m})=\frac{1}{2}\sum_{j=1}^3 |n_j-m_j|  \,,
\end{equation}
i.e.\ the minimal number of photons that have to be moved in order to turn
$\boldsymbol{n}$ into $\boldsymbol{m}$.

The quantum mechanical distance between the evolving and the initial state has
to take into account their probabilities on the Fock states. These probabilities
are formally defined in \eqref{eq:FSLamplitude}-\eqref{eq:FSLentropy} in the
main text; to simplify the notation, here we denote them by
$p_{\boldsymbol{n}}^{(\sigma)}(t)$ and $p_{\boldsymbol{m}}^{(\sigma)}(0)$ for
the evolving and the initial state respectively.
To account for them in the quantum graph distance, we introduce a weight
$F_{\boldsymbol{n} \boldsymbol{m}}(t)$ associated with each pair of vertices.
These weights play the role of a joint probability distribution whose marginals
are $p_{\boldsymbol{n}}^{(\sigma)}(t)$ and $p_{\boldsymbol{m}}^{(\sigma)}(0)$,
and we therefore impose the constraints
\begin{equation}
    \sum_{\boldsymbol{m}\in {\rm FSL}}F_{\boldsymbol{n} \boldsymbol{m}}(t)=p_{\boldsymbol{n}}^{(\sigma)}(t)\,,
    \qquad
     \sum_{\boldsymbol{n}\in {\rm FSL}}F_{\boldsymbol{n} \boldsymbol{m}}(t)= p_{\boldsymbol{m}}^{(\sigma)}(0)
\label{eq:weight_constraint}
    \,,
\end{equation}
where the sums run over all the vertices of the FSL.
The weights $F_{\boldsymbol{n} \boldsymbol{m}}$ are not fixed by first
principles: the constraints \eqref{eq:weight_constraint} leave a whole family of
admissible choices. It is then natural to select the cheapest one, defining
\begin{equation}
\label{eq:dG_Wasserstein}
d^{(W)}_\mathrm{G}(t)=\min_{F\geq 0} \sum_{\boldsymbol{n},\boldsymbol{m}\in {\rm FSL}} F_{\boldsymbol{n} \boldsymbol{m}}(t)
d(\boldsymbol{n},\boldsymbol{m})\,,
\end{equation}
where the minimization is performed over all the weight functions
$F_{\boldsymbol{n} \boldsymbol{m}}$ satisfying \eqref{eq:weight_constraint}.
The definition \eqref{eq:dG_Wasserstein} is the Wasserstein-1
distance \cite{Villani2009}, which describes the cost of the optimal transport on the FSL and justifies the the superscript $(W)$ used to label this quantity. From a physical point of view, it corresponds to the minimum probability-weighted
number of photon moves needed to turn the initial distribution into the one at time $t$. The minimization is also what guarantees that
$d^{(W)}_\mathrm{G}(t)$ is a genuine distance, in particular that it satisfies the
triangle inequality.
At the level of quantum states it remains a
pseudo-distance, since it depends on $|\psi(t)\rangle$ only through the
populations $p^{(\sigma)}_{\boldsymbol{n} }(t)$ and is therefore blind to their relative
phases.

Although the evaluation of \eqref{eq:dG_Wasserstein} requires solving an
optimization problem, it has the advantage of being valid for any initial state
in a given charge sector of the model, not necessarily belonging to the Fock
basis. Moreover, for specific choices of the initial state it reduces to more
easily computable expressions, and for the case considered in
Sec.\,\ref{subsec:graphdistance} it reduces to \eqref{eq:graphdistance}.
Indeed, if the initial state is a Fock state labeled by the vector of
occupation numbers $\boldsymbol{v}$, i.e.\ $p^{(\sigma)}_{\boldsymbol{m}}(0)=\delta_{\boldsymbol{m},\boldsymbol{v}}$,
the second constraint in \eqref{eq:weight_constraint} together with the
non-negativity of $F_{\boldsymbol{n} \boldsymbol{m}}$ implies
$F_{\boldsymbol{n} \boldsymbol{m}}(t)=0$ whenever $\boldsymbol{m}\neq \boldsymbol{v}$.
The first constraint then gives
$F_{\boldsymbol{n} \boldsymbol{v}}(t)=p_{\boldsymbol{n}}^{(\sigma)}(t)$, so that
the admissible weight is unique, and
\begin{equation}
\label{eq:graph distance Fock}
    d^{(W)}_\mathrm{G}(t)= \sum_{\boldsymbol{n}\in {\rm FSL}} p_{\boldsymbol{n}}^{(\sigma)}(t)
d(\boldsymbol{n},\boldsymbol{v})\,.
\end{equation}
If in addition the initial state represents a corner of the FSL, i.e.\
$\boldsymbol{v}=\mathcal{N}_3 \boldsymbol{e}_j$, where $\boldsymbol{e}_j$ is one of the
three elements of the canonical basis of a three-dimensional vector space, then
\begin{equation}
\label{eq:graph distance corner}
    d(\boldsymbol{n},\mathcal{N}_3 \boldsymbol{e}_j)=\frac{1}{2}\left(\mathcal{N}_3-n_j+\sum_{i\neq j}n_i\right)=\mathcal{N}_3-n_j\,,
\end{equation}
where we have used the
conservation of the photon number in each charge sector in the last equality.
Plugging \eqref{eq:graph distance corner} into
\eqref{eq:graph distance Fock}, we retrieve \eqref{eq:graphdistance} (adapted to
a generic FSL corner initial state).

We conclude by mentioning that the same construction can be carried out away
from the large-detuning regime, where the FSL is the one shown, for the charge
sector $q_3=2$, in the left panel of Fig.\,\ref{fig:JC_N3_FSLs}. Two things change in this case. First, every FSL
vertex is labeled not only by the three bosonic occupation numbers, but also by
the atomic excitation, i.e. the eigenvalue of $(\sigma^{(z)}+1)/2$. Second, the
classical graph distance itself is modified: on the hexagonal FSL a photon
cannot hop directly between two modes, but has to be absorbed and re-emitted by
the spin, so that each mode-to-mode transfer costs two steps and
$d(\boldsymbol{n},\boldsymbol{m})=\sum_{i=1}^{3}|n_i-m_i|$. With this
replacement, \eqref{eq:dG_Wasserstein} remains the same and provides the
required generalization, which, when the initial state is among
$|q_3,0,0;-1\rangle$, $|0,q_3,0;-1\rangle$ and $|0,0,q_3;-1\rangle$, reduces to
\eqref{eq:graphdistance_2ndexample}.

A similar procedure can be applied to the JC model with a generic number of
modes, both in and away from the large-detuning limit. A complete and general
treatment is beyond the scope of this work, and we hope to come back to this
problem in future investigations.

The same construction can also be carried out for the
Dicke model of Sec.\,\ref{Dicke}. Equation \eqref{eq:dG_Wasserstein} applies to
that case as well, provided one uses the appropriate classical graph distance,
which differs between the resonant and the off-resonant regimes because, in the
polariton basis, the corresponding FSLs are different. At
resonance, if the initial state is the polariton vacuum, i.e.\ the origin of the
FSL, \eqref{eq:graphdista_rect} is retrieved.

\section{Derivation of the Dicke graph distance off-resonance}
\label{app:off-resonance}

In this appendix we report the derivation of \eqref{eq:offi diag dist} in the main text. We begin from \eqref{HP Hamiltonian} and we define  the canonical quadratures $x_a, \,p_a$ and $x_b,\, p_b$ through 
\be
a=\sqrt{\frac{\omega}{2}} (x_a+ \frac{i}{\omega}p_a), \quad b=\sqrt{\frac{\Delta}{2}} (x_b+ \frac{i}{\Delta}p_b)~.
\ee
In terms of these variables, the Hamiltonian assumes the form
\be
H_{\rm HP}=\frac{1}{2}(p_a^2 +\omega^2x_a^2)+\frac{1}{2}(p_b^2 +\Delta^2x_b^2)
+2g\sqrt{\omega \Delta} x_ax_b~,
\ee
up to additive constants.
This is the textbook example of two coupled oscillators with unit mass matrix and potential matrix
\be
V=\begin{psmallmatrix}\omega^2&2g\sqrt{\omega \Delta}\\ 2g\sqrt{\omega \Delta}&\Delta^2\end{psmallmatrix}~.
\ee
\\
The matrix $V$ is diagonalized by the rotation
$R(\theta)=\left(\begin{smallmatrix}\cos\theta&&-\sin\theta\\
\sin\theta&&\cos\theta\end{smallmatrix}\right)$, with
\begin{equation}
\cos2\theta=\frac{\omega^2-\Delta^2}{\Xi}\,,\qquad
\sin2\theta=\frac{4g\sqrt{\omega\Delta}}{\Xi}\,,
\end{equation}
where $\Xi=\sqrt{(\omega^2-\Delta^2)^2+16g^2\omega\Delta}$ is the quantity
introduced above \eqref{eq:offi diag dist}. Its eigenvalues are the squared
normal-mode frequencies
\begin{equation}
\label{eq:alpha_pm_app}
\alpha^2_\pm=\frac{\omega^2+\Delta^2}{2}\pm\frac{\Xi}{2}\,,
\end{equation}
so that $\Xi=\alpha^2_+-\alpha^2_-$, as in the main text.

This allows us to define normal coordinates in terms of the quadratures $\mathbf{X}=R^\top \mathbf{x}$ and $\mathbf{P}=R^\top \mathbf{p}$ which decouple the Hamiltonian leading to 
\be
\label{eq:Ham_Dicke_quadratures}
H=\frac{1}{2} ( P_+^2+\alpha_+^2X_+^2 )+\frac{1}{2} ( P_-^2+\alpha_-^2X_-^2 )~.
\ee
From this form it is easier to describe the time evolution of the normal-mode quadratures, with the Heisenberg equations leading to the solutions 
\be
X_\pm(t)=\cos{(\alpha_\pm t)}X_\pm +\frac{\sin{(\alpha_\pm t)}}{\alpha_\pm}P_\pm, \quad P_\pm(t)= -\alpha_\pm \sin{(\alpha_\pm t)} X_\pm +\cos{(\alpha_\pm t)}P_\pm ~,
\ee
where, in the rest of the appendix, the quadratures without argument are operators at time $t=0$.
At this point, we notice that the transformation $c_\pm=\frac{a\pm b}{\sqrt{2}}$ leaves the total number operator invariant. Using this identity, \eqref{eq:graphdista_rect} can
equivalently be written, switching to the Heisenberg representation, as
\begin{equation}
\label{eq:graphdistance_a_b}
 d_\mathrm{G}(t)= \frac12\bra{\psi_0}a(t)^\dagger a(t)\ket{\psi_0}
 +\frac12\bra{\psi_0}b(t)^\dagger b(t)\ket{\psi_0}\,,
\end{equation} 
where $\ket{\psi_0}$ is the polariton vacuum considered also in the main text.
The graph distance in \eqref{eq:graphdistance_a_b} is expressed in terms of the expectation values 
\be \label{bare distance}
d_{\rm G}(t)= \frac{1}{4}\left(\frac{\braket{p_a^2}}{\omega} +\omega \braket{x_a^2}-1\right) +\frac{1}{4}\left(\frac{\braket{p_b^2}}{\Delta} +\Delta \braket{x_b^2}-1\right)~,
\ee
where here, to lighten the notation, we are denoting $\langle\cdot\rangle\equiv \bra{\psi_0}\cdot\ket{\psi_0}$.
After implementing the rotation via $R(\theta)$,
in the rotated coordinates we compute
\begin{align}
    &\braket{X_\pm (t)^2}=\cos^2{(\alpha_\pm t)}\braket{X_\pm^2} +\frac{\sin^2{(\alpha_\pm t)}}{\alpha_\pm^2}\braket{P_\pm^2}\,,\\ &\braket{P_\pm(t)^2}= \alpha_\pm^2 \sin^2{(\alpha_\pm t)} \braket{X_\pm^2} +\cos^2{(\alpha_\pm t)}\braket{P_\pm^2}\,,\\
    &\braket{X_+(t)X_-(t)}=\frac{g \sqrt{\omega \Delta}(\omega-\Delta)}{\Xi} \left( \frac{\cos{(\alpha_+t)}\cos{(\alpha_-t)}}{\omega \Delta} -\frac{\sin{(\alpha_+t)}\sin{(\alpha_-t)}}{\alpha_+\alpha_-}\right)\,,\\ 
    & \braket{P_+(t)P_-(t)}=\frac{g \sqrt{\omega \Delta}(\omega-\Delta)}{\Xi} \left(\frac{\alpha_+\alpha_-}{\omega \Delta}\sin{(\alpha_+t)}\sin{(\alpha_-t)}   -\cos{(\alpha_+t)}\cos{(\alpha_-t)}\right)\,,
\end{align}
where we have used that $\braket{X_iP_j +P_jX_i}=0$.
Substitution of these expressions in \eqref{bare distance} leads to
\be
\begin{split}
    d_{\rm G}(t)=&\frac{2g^4(\omega +\Delta)^2}{\Xi^2}\left(\frac{\sin^2{(\alpha_+ t)}}
    {\alpha_+^2}+\frac{\sin^2{(\alpha_- t)}}{\alpha_-^2}\right)+\\
    &\frac{2g^2(\omega -\Delta)^2}{\Xi^2}\left(1-\cos{(\alpha_+t)} \cos{(\alpha_-t)}+\frac{\omega\Delta -2g^2}  {\alpha_+\alpha_-}\sin{(\alpha_+t)} \sin{(\alpha_-t)}\right)~,
\end{split}
\ee
which is the one appearing in \eqref{eq:offi diag dist} in the main text.

\bibliographystyle{JHEP}
\bibliography{Refs.bib}

\end{document}